\documentclass[12pt]{article}

\usepackage{graphicx}
\usepackage{caption}
\usepackage{times}
\usepackage{lineno}
\usepackage{epstopdf}

\newenvironment{sciabstract}{%
\begin{quote} \bf}
{\end{quote}}

\newcounter{lastnote}

\title{Polarimetric evidence of a white dwarf pulsar in the binary system AR Scorpii}

\author
{D.A.H. Buckley$^{1*}$, P.J. Meintjes$^{2}$, S.B. Potter$^{1}$, T. R. Marsh$^{3}$\\
and B.T. G\"{a}nsicke$^{3}
$\\
\\
\normalsize{$^{1}$  South African Astronomical Observatory, PO Box 9, Observatory, 7935, Cape Town, South Africa}\\
\normalsize{$^{2}$ Department of Physics, University of the Free State, PO Box 339, Bloemfontein, 9300,  South Africa}\\
\normalsize{$^{3}$ Department of Physics, Gibbet Hill Road,  University of Warwick, Coventry, CV47AL, UK }\\
\\
\normalsize{$^{*}$ E-mail: dibnob@saao.ac.za}
}

\date{}

\begin{document} 


%\baselineskip24pt


\maketitle


\begin{sciabstract}
The variable star AR Sco was recently discovered to pulse in brightness every 1.97 min from ultraviolet wavelengths into the radio regime. The system is composed of a cool, low-mass star in a tight, 3.55 hr orbit with a more massive white dwarf. Here we report new optical observations of AR Sco that show strong linear polarization (up to 40\%) which varies strongly and periodically on both the spin period of the white dwarf and the beat period between the spin and orbital period, as well as low level ($<$ a few \%) circular polarization. These observations support the notion that, similar to neutron star pulsars, the pulsed luminosity of AR Sco is powered by the spin-down of the rapidly-rotating white dwarf which is highly magnetised (up to 500 MG). The morphology of the modulated linear polarization is similar  to that seen in the Crab pulsar, albeit with a more complex waveform owing to the presence of two periodic signals of similar frequency. Magnetic interactions between the two component stars, coupled with synchrotron radiation from the white dwarf, power the observed  polarized  and non-polarized emission. AR  Scorpii is therefore the first example of a white dwarf pulsar.
    
\end{sciabstract}

\section*{Background}

Radio pulsars, discovered nearly 50 years ago, are fast rotating magnetised neutron stars with spin-modulated synchrotron radio emission, powered by spin-down energy loss of fast rotating neutron stars ({\it 1\/}).  The star AR Scorpii was recently discovered to be a 3.56 h close binary, containing a fast rotating (117 s) white dwarf, showing strong brightness variations across most of the electromagnetic spectrum (ultraviolet to radio), most strongly modulated on the 118.2 s beat (synodic) period, and its harmonics ({\it 2\/}).  The spin-down of the white dwarf ($\dot{P} = 3.92 \times 10^{-13}  \, \mbox{s s}^{-1}$) powers non-thermal emission, whose luminosity far exceeds  (by a factor of ${\geq}$14; {\it 2}) the combined luminosity of the stellar components and dominates the spectral energy distribution. These observations were explained in terms of beamed synchrotron radiation from the white dwarf, some of which is reprocessed by the companion star  ({\it 2}).  The weak X-ray emission suggests that little accretion power is produced in AR Sco, which either implies that it is currently in a propeller mass ejection phase  or there is no mass transfer at all. If the former, then it would be similar to the white dwarf in the cataclysmic variable AE Aquarii ({\it 3, 4, 5, 6\/)}, which has a 33 s spin period (s) and a $\dot {P} = 5.6 \times 10^{-14} \, \mbox{s s}^{-1}$. However, the lack of flickering and broad emission lines in AR Sco, indicative of mass outflows which are seen in AE Aqr, implies no mass loss and suggests that a different mechanism is draining the rotational kinetic energy from the rapidly rotating white dwarf in AR Sco, perhaps similar to that operating in pulsars, namely dipole radiation (e.g. {\it 4}) and MHD interactions.

\section*{Results} 
We present and explain the first polarimetric study of AR Sco, which shows strong pulsed linear polarization (up to 40\%), highly modulated on the spin and beat periods, and little circular polarization (a few percent at most). We develop a model which interprets the modulated polarized emission, as well as the earlier observations ({\it 2\/}) of the unpolarized emission, in terms of  synchrotron radiation produced at two separate sites: the magnetosphere of the magnetic white dwarf, and interaction regions involving the magnetosphere of the M-star. The combined synchrotron emission from these sites produces the pulsed emission seen at both the spin and beat periods, as well as producing a synchrotron dominated SED.

\subsection*{Photopolarimetry}
High speed all-Stokes optical polarimetry of AR Sco was obtained using the HIPPO polarimeter ({\it 7\/)} on the SAAO 1.9 m telescope on two consecutive nights in March 2016 in two broad spectral bands (see Methods: Data sources). These observations revealed the system to be strongly linearly polarized, reaching levels of 40\%. This is demonstrated in Figure 1, where the flux and polarization variations in the broad red band on the first night are presented. We also show additional observations and variations of the Stokes parameters ({\it I, Q, U, V}) on both nights in the Methods section. Obvious are the strongly modulated total intensity variations, found from a period analysis discussed below, to be predominantly at the beat period and its harmonics, as previously reported {\it (2\/)}. The degree of linear polarization, {\it p}, shows pulse fractions up to 90\%, while the position angle is seen to rotate through $\sim 180^{\circ}$ at the frequency of the first harmonic of the spin or beat frequency. In contrast, the level of circular polarization is comparatively low, at a few percent, with variations in {\it V} that are sometimes seen to be correlated with the {\it Q, U} variations, when they are most strongly modulated.

\subsection*{Periodicities}
The data were subjected to a period analysis (see Methods) and the power spectra for the Stokes {\it Q} and {\it U} parameters are presented in Figure 2. We see power at both the spin ($\omega$) and beat ($\omega - \Omega$) frequencies (where $\omega$ and $\Omega$ are the spin and orbital frequencies, respectively). These results show that, as expected for a dipolar white dwarf magnetic field, the modulation is strongest at the first harmonic of the spin period, namely 58.55 s, implying that polarized emission is seen from both magnetic poles. These results can be explained by the white dwarf magnetic field producing a magnetically-confined bipolar outflow of charged particles emitting strong linearly polarized synchrotron radiation which, together with the magnetic field of the white dwarf, sweeps across our line of sight and across the inner face of the M-star companion, where it leads to unpolarized reprocessed optical emission. 

\subsection*{Spin modulated polarization} 
The spin phase-folded polarimetry is shown in Figure 3, which clearly shows the double-peaked variations of the polarized intensity,  and particularly the strongly modulated ($\sim$90\% pulse fraction) linear polarization, together with the large swing in position angle through $180^\circ$, a likely consequence of viewing the dipole perpendicular to its axis. The folded data are different on the two nights, with a much higher percentage polarization and a higher pulse fraction seen on the first night. The different nature of the spin modulations on the two nights maybe due to the different orbital phases of the two observations, which were at $\phi_{orb} \sim$ 0.15 and $\phi_{orb} \sim$ 0.64, respectively, though further observations over a number of orbital cycles will be needed to confirm this. Both the spin and beat modulations combine differently on the two nights, resulting in the variability of the waveforms for the phase-folded variations. Extracting the two pure spin and beat and signals, as a function of orbital phase, will require more extensive observations. Likewise for discriminating between repeatable orbital modulations and stochastic effects.

In Figure 4 we plot the average {\it Q} and {\it U} data pairs over a white dwarf spin cycle and draw the trajectories of their motion in the {\it Q$-$U} plane, which follow counter-clockwise loops and demonstrate how they vary as the magnetic orientation changes. On the first night ($\phi_{orb} \sim$ 0.15), the main peak maps to the outer circle and the secondary peak maps to the lower half of the small loop inside it. For the observation on the second night ($\phi_{orb} \sim $ 0.65), there is an apparent phase change and more complex polarized flux variations, leading to a different trajectory in the {\it Q$-$U} plane, possibly indicative of a variation in the magnetic field orientation over the two nights due to orbital modulation. 

These {\it Q} and {\it U} variations are qualitatively similar to that seen in the optical polarimetry of the Crab pulsar ({\it 8\/)}, although the latter's {\it p} and $\theta$ variations show rather different morphologies, with more abrupt $\theta$ ``swings" than we see for AR Sco in our lower resolution data. While the polarized flux was not shown for the Crab in this study, the duty cycle of the light curve for the Crab is $\sim$ 30\% (for a normalized intensity ${\ge} 0.1$), much less that the minimum duty cycle of $\sim 60\%$ seen for the polarized and total UV flux ({\it 2})\/ in AR Sco.  Also, there are a range of different $\theta$ swing morphologies observed in radio pulsars ({\it 9\/}), which have been interpreted in terms of a rotating vector model (RVM; {\it 10\/)}, where the linear polarization vector ({\it p}, $\theta$) represents a projection of the magnetic field where the polarized radiation is produced on the plane of the sky. In this model the emission is confined to the magnetic poles, where the field lines of the dipole are essentially parallel to each other, with a small opening angle. The morphologies of {\it Q--U} ``loop diagrams" have been used to model other magnetic systems, for example the oblique rotator model as applied to the polarized Zeeman spectra of magnetic Ap stars ({\it 11\/}).

The polarization behaviour of AR Sco is quite different from the other polarized asynchronous magnetic white dwarf binaries, namely intermediate polars (IPs) which, unlike AR Sco, are accretion luminosity dominated systems, with significant X-ray emission. In the nine IPs where polarization has been measured, it is typically only at the few percent level ({\it 12, 13\/}), and is predominantly circularly polarized, as expected from cyclotron emission from an accreting magnetic white dwarf. The high level of linear polarization and lower level of circular polarization is consistent with synchrotron emission, where a maximum value for the latter of $\sim 15\%$ is expected ({\it 14\/}).   

\section*{Constraining the white dwarf magnetic field}

The lack of the usual signatures for accretion, namely flickering and strong, broad and variable Doppler broadened emission lines, coupled with the relatively low X-ray luminosity of $\sim 5 \times 10^{30}  \, \mbox{erg s}^{-1}$ ({\it 2\/}), some two orders of magnitude less than a typical accreting magnetic white dwarf, is evidence that the white dwarf in AR Sco is currently not accreting, or has a very low accretion rate. The ratio of X-ray to spin-powered luminosity ($\alpha = 3.3 \times 10^{-3}$; see Methods: Luminosities) indicates that the white dwarf in AR Sco is currently behaving as a rotation powered pulsar, similar to AE Aqr ({\it 4}; and Supplementary Figure 8). Most of the luminosity of the system is being driven by rotational kinetic energy losses. The bulk of the spin-down power is expected to be lost by a combination of magnetic dipole  radiation (see Methods: Magnetic dipole radiation and ({\it 4}; and references therein), as well as magnetohydrodynamic (MHD) interactions of the fast rotating white dwarf's magnetic field with the secondary star. Additional losses may also come from an out-flow of relativistic charged particles from the magnetic white dwarf and a wind from the M-star companion. 

The upper limit on the magnetic dipole strength can be derived assuming that the bulk of the spin-down power is radiated by dipole radiation (Poynting radiation; see Methods: Magnetic dipole radiation). So if $L_{\rm m-d} = L_{\dot{\nu_{\rm s}}}$, the upper limit placed on the white dwarf magnetic field for a maximum dipole tilt angle to the spin axis of $\chi = 90^{\circ}$, is given as
\begin{eqnarray}
B_{1, *} & \approx & 500 \, \left(\frac{L_{\dot{\nu_{\rm s}}}}{1.5 \times 10^{33} \, \mbox{erg s}^{-1}} \right)^{1/2} \left(\frac{P_{\rm wd}}{117 \, \mbox{s}}\right)^{2} 
\left(\frac{R_{\rm wd}}{5.5 \times 10^{8} \, \mbox{cm}}\right)^{-3} \, \mbox{MG}. 
\end{eqnarray}
This value is in the regime for high field magnetic white dwarfs, either isolated or in magnetic cataclysmic variables ({\it 15}). 

Another approach, assuming that a fraction of the spin-down power is dissipated through a magnetic stand-off shock near the secondary,  gave an estimate of the magnetic field of $\sim$100 MG ({\it 16}). If rotational energy is also dissipated through magnetohydrodynamical (MHD) pumping of the secondary star, then a constraint can be placed by estimating the MHD power dissipated in the surface layers of the secondary ({\it 17}). Dissipation will occur through magnetic reconnection and Ohmic heating, particularly in that part of the secondary star's photosphere which faces the white dwarf. This could contribute to both the observed line emission and the strong orbital photometric modulation, which is at maximum (after removal of the spin pulsations at ${\phi} _{orb} \sim 0.5$ ({\it 2}), when the secondary star is at superior conjunction.   

MHD pumping of the secondary allows setting an independent constraint on the white dwarf surface magnetic field strength. If the white dwarf magnetic field lines sweep periodically across the secondary star, the penetration depth of the magnetic flux into the surface layers is (see e.g. {\it 17} and references therein)  $\delta=\sqrt{2 \eta_{\rm tur}/\omega_{\rm b}}$ (Methods: Spin-down torque), where $\eta_{\rm tur}$ and $\omega_{\rm b}$ are the turbulent diffusivity and beat (synodic) angular frequency, respectively. For the photospheric conditions of a M-type dwarf,   $\delta \sim 10^{8} \, \mbox{cm}$, i.e. using $\eta_{\rm tur} \leq 10^{15} \, \mbox{cm}^{2}\mbox{s}^{-1}$ ({\it 17}). The power dissipation of magnetic energy through reconnection and Ohmic heating can be estimated from $P_{\rm mhd} = (B^{2}/8 \pi) (4 \pi R_{2}^{2} \delta)~\omega_{\rm b}$ (e.g. {\it 17} and references therein).

For a 500 MG magnetic white dwarf, the field strength at the distance of the secondary (at a distance of $a = 8 \times 10^{10} \, \mbox{cm}$; see Methods: Binary parameters) will be 160 G. This implies $P_{\rm mhd} = 4 \times 10^{31} \, \mbox{erg s}^{-1}$, which is $\sim 30\%$ of the average optical luminosity of AR Sco in excess of the combined stellar contributions, namely $L_{+} = 1.3   \times 10^{32} \, \mbox{erg s}^{-1}$ ({\it 2}). The equation for power dissipation through MHD pumping can then be expressed as:

\begin{eqnarray}
%B & = & \left(\frac{8 \pi P_{\rm mhd}}{4 \pi R_{2}^{2} \delta \omega_{\rm b}}\right)^{1/2} \nonumber \\
B  & \sim & 160 \left(\frac{P_{\rm mhd}}{4 \times 10^{31} \, \mbox{erg s}^{-1}} \right)^{1/2} \left(\frac{R_{2}}{2.5 \times 10^{10} \, \mbox{cm}}\right)^{-1} \left(\frac{\delta}{10^{8} \, \mbox{cm}}\right)^{-1/2} \left(\frac{P_{\rm b}}{118 \, \mbox{s}}\right)^{1/2} \, \mbox{G},
\end{eqnarray}

\noindent implying a dipole surface  magnetic field of
\begin{eqnarray}
 B_{1, *} = B \, (a/R_{\rm wd})^{3} \leq 500  (B/160 \, \mbox{G}) (a/8 \times 10^{10} \, \mbox{cm})^{3} (R_{\rm wd}/5.5 \times 10^{8} \, \mbox{cm})^{-3} \, \mbox{MG}, 
\end{eqnarray}

\noindent
where it was assumed that the secondary star fills, or nearly fills, its Roche lobe, so equals $R_{\rm L,2}$ (Methods: Binary parameters). If other processes are also responsible for the line emission and heating of the secondary, for example beamed synchrotron radiation, dipole radiation or charged particles, all from the white dwarf, then this will lower $P_{\rm mhd}$ and hence the estimate of {\it B} and $B_{1,*}$ based on MHD interactions alone. 

\section*{The white dwarf pulsar}
We propose that the highly asynchronous binary system AR Sco contains a strongly magnetic white dwarf, probably having been spun-up to the current short rotation period ($P_s $= 117 s) by accretion torques during a high mass transfer phase in its history, as has also been proposed for AE Aqr (e.g. {\it 4,5,6\/}). The observed ratios of X-ray to spin-power luminosities for AR Sco and AE Aqr  ({\it 2,4}) are very similar to spin-powered neutron star pulsars (e.g. {\it 18\/}; and Supplementary Figure 8), making both of them analogous to spun-up radio pulsars ({\it 4, 19}). Similarly,  considering that the 70\% pulse fraction of the luminosity in excess of the combined stellar components ({\it 2}) is predominantly optical synchrotron emission, this implies that the synchrotron power in AR Sco is $0.06 P_{s-d}$,  similar to the ratio of synchrotron produced gamma-ray emission to spin-down power reported recently for a sample of spun-up gamma-ray emitting milli-second radio pulsars ({\it 20}). 

The high level of linear polarization in AR Sco is consistent with synchrotron emission of relativistic electrons in ordered magnetic fields (e.g. {\it 14}). The periodicities, at both the white dwarf spin and the beat periods, are also consistent with this emission being produced in the white dwarf magnetosphere, which is additionally modulated at the binary period. 

The spectral energy distribution (SED) in AR Sco ({\it 2, 21}) shows a $ S_{\nu} \propto \nu^{\alpha_{1}}$ ($\alpha_{1} \sim 1.3$) self-absorbed power law spectral distribution for $\nu \leq 10^{12} - 10^{13} \, \mbox{Hz}$, i.e. at infrared to radio wavelengths. We suggest that these originate from pumped coronal loops of the nearly Roche lobe-filling secondary star (e.g. {\it 22,23, 24}). The magnetospheric flux tubes of the secondary star are distorted by the fast rotating white dwarf dipolar field (see Methods: Magnetic interactions), inducing strong field-aligned potentials (e.g. {\it 25, 26}), resulting in synchrotron flares through the van der Laan process ({\it 5, 27, 28\/}), whose superposition produces the observed power law at $\nu \leq 10^{12} \, \mbox{Hz}$ and the peak emission at $\nu_{\rm crit} \sim 0.3 \nu_{\rm syn} \leq 10^{13} \, \mbox{Hz}$. This  emission is expected to be pulsed at the beat frequency, consistent with the ATCA observations at 5.5 and 9.0 GHz ({\it 2\/}), and may be linearly and/or circularly polarized, depending on the orientation of the flux tubes relative to the observer. 

At higher frequencies, $\nu \geq \mbox{few} \times 10^{14} \, \mbox{Hz}$ (optical-UV-X-rays), the SED follows a different $\nu^{\alpha_{2}}$ power law ({\it 2}), clearly distinguishable from the spectral components of both the M5 secondary and the white dwarf, where $\alpha_{2} \sim -0.2$ (also see {\it 21}). This component is produced by non-thermal synchrotron emission from the magnetic white dwarf dipole and shows a high level of linear polarization.

The absence of accretion, and hence conducting plasma, allows for the induction of significant electrical potentials ({\it 29}) of the order of $\Delta V \sim 10^{12} \, \mbox{V}$ in the vicinity of the light cylinder, $r_{\rm l-c} \sim 6 \times 10^{11} \, (\omega_{\rm wd}/0.054 \, \mbox{rad s}^{-1}) \, \mbox{cm}$, where the white dwarf magnetic field is of the order $B_{\rm l-c} \sim 0.4 \, \mbox{G}$, which is $\sim 7.5 \times$ greater than the orbital separation (see Methods: Binary parameters). This electric field can produce a relativistic wind of electrons (and ions), with energies of the order of $\gamma_{\rm e} \sim 10^{6}$,  which will emit synchrotron radiation at  frequencies up to $\nu_{\rm syn} \leq 3 \times 10^{17} (B_{\rm l-c}/0.4 \, \rm G) (\gamma_{\rm e}/10^{6})^{2} \, \mbox{Hz}$ (soft X-rays) in the vicinity of the light cylinder radius. Since the whole binary system is inside the accelerator zone, i.e. the light cylinder radius (see Methods: Binary parameters), this emission may be modulated at both the spin and beat periods.

If the white dwarf is an oblique rotator, with the magnetic dipole tilted to the spin axis, the rotating magnetosphere is expected to produce a relativistic outflowing MHD wind outside the light cylinder, with alternating regions of opposite polarity, i.e. a ``striped" wind (see e.g. {\it 30, 31, 32\/}; Methods: Pulsar-like particle acceleration), resulting in additional particle acceleration through magnetic reconnection and associated pulsed incoherent polarized synchrotron emission ({\it 30, 31}).  The emission will be pulsed at the spin period (fundamental and first harmonic, arising from both magnetic poles), as well as displaying the 180$^{\circ}$ swing of the position angle of the polarization (see e.g. {\it 32} for a discussion). The binary motion of the system may also impart a beat period modulation, seen in both the polarized and unpolarized optical emission.

\section*{Discussion}

The high degree of asynchronism in AR Sco ($P_s / P_{orb} = 0.009$) indicates a previous spin-up phase of the white dwarf, due to a higher mass transfer rate,  similar to other highly asynchronous intermediate polar magnetic cataclysmic variables, like AE Aqr, DQ Her, XY Ari and GK Per (see e.g. {\it 33\/}). However, these system are still undergoing mass transfer but are now likely to be in, or close, to spin equilibrium. For AE Aqr it has been shown ({\it 34, 35}) that during a high mass transfer phase the secondary star may have shed its outer envelope in a catastrophic run-away mass transfer process, resulting in the white dwarf being spun-up to a period close to its current value of $\sim 33 s$.  There is currently no observational evidence that AR Sco has evolved through such an extreme high mass transfer phase and so the evolutionary path which AR Sco took to its current configuration is still therefore an open question. 

Given the high magnetic fields we are estimating, AR Sco presents a significant problem when it comes to spinning up the white dwarf in the first place, since material will tend to be ejected rather than accreted, except at very high accretion rates. Since the observed white dwarf spin-down timescale of $\sim 10^{7} \, \mbox{yr}$ ({\it 2}) is less than the spin-orbit synchronization timescale of $\sim 2.5 \times 10^{8} \, \mbox{yr}$, calculated for MHD torques alone (Methods: Spin-down torque), this implies the bulk of the spin-down power is dissipated through magnetic dipole radiation and other channels. The high value of the magnetic field is therefore the reason for the current large spin down rate, through the various mechanisms (e.g. dipole radiation, MHD interactions) we have described here, which rob the white dwarf of its angular momentum. 

The strongly pulsed polarized optical emission in AR Sco is analogous to that observed in pulsars, like the Crab. The ratio $\alpha = (L_{\rm x}/L_{\rm s-d}) \sim 10^{-3}$ derived for AR Sco implies that most of the luminosity of the system is not produced by accretion of matter, but by spin-down energy loss, implying that the white dwarf behaves like a spin-down powered pulsar (Supplementary Figure 8). Electric potentials of the order $ \Delta V  \sim  10^{12} \, \mbox{V}$ can be induced between the white dwarf and the light cylinder (Methods: Magnetic field interactions), accelerating particles to energies of the order of $\gamma_{\rm e} \sim 10^{6}$ (Methods: Pulsar-like particle acceleration), resulting in pulsed and strongly polarized synchrotron emission, possibly up to to  X-ray frequencies. However, we argue that a striped relativistic magnetohydrodynamic wind, outside the light cylinder, may also be present (Methods:Pulsar-like particle acceleration). The pulsed emission below optical frequencies (i.e. radio) arises from synchrotron emission in pumped coronal loops of the secondary star, which is consistent with the observed SED ({\it 2, 21}). Future observations, particularly at X-ray and radio wavelengths, will be important in determining the exact nature of the emission mechanisms operating in AR Sco. More extensive time resolved polarimetry will also help to disentangle the two closely spaced polarized signals, at the spin and beat period, leading to more definitive conclusions regarding geometry.

\noindent
\newline
{\bf Contact author}
All correspondence regard this paper and information regarding access to the data and materials presented in this paper should be addressed to the lead author (DAHB; dibnob@saao.ac.za).

\noindent
\newline
{\bf Acknowledgments}
For DAHB, PJM and SBP, this work was supported by the National Research Foundation of South Africa. TRM was supported by the Science and Technology Facilities Council (STFC) under grant ST/L000733. BTG is supported through European Research Council grant 320964. This work is based on observations obtained at the South African Astronomical Observatory.

\noindent
\newline
{\bf Author roles}
DAHB conceived of the HIPPO observing programme and organized the observations, assisted in the analysis and interpretation of the polarimetry, participated in the modelling and was primary author of the paper. PJM undertook the modelling and led most of the interpretation.  SBP undertook the reductions of the HIPPO data, produced most of the figures and assisted in interpretation of the results. TRM and BTG provided information on AR Sco, including pre-publication material, and assisted in the interpretation of the results and models.

\noindent
\newline
{\bf Data Availability}
The data used to produce the plots in this paper and other findings of this study are available from the corresponding author upon reasonable request. All of the reduced HIPPO photopolarimetry is available on the SAAO CloudCape website for downloading at:\\
https://cloudcape.saao.ac.za/index.php/s/npMF46F1K56fsPt.

\noindent
\begin{quote}
\section*{References}

\noindent
\begin{enumerate}
\item T. Gold, Rotating neutron stars as the origin of the pulsating radio sources. {\it Nature \/} {\bf 218}, 731-732 (1968).
\item T.R. Marsh et.al., A radio-pulsing white dwarf binary star. {\it Nature \/} {\bf 537}, 374-377 (2016)
\item G.A. Wynn, A.R. King, K. Horne, A magnetic propeller in the cataclysmic variable AE Aquarii. {\it Mon. Not. R. Astron. Soc.\/} {\bf 286,} 436-446 (1997)
\item N.R. Ikhsanov, The pulsar-like white dwarf in AE Aqarii. {\it Astron. Astrophys. \/} {\bf 338}, 521-526 (1998)
\item P.J. Meintjes, L.A. Venter, The diamagnetic blob propeller in AE Aquarii and non-thermal radio to mid-infrared emission. {\it Mon. Not. R. Astron. Soc.\/} {\bf 360}, 573-582 (2005)
\item B. Oruru, P.J. Meintjes, X-ray characteristics and the spectral energy distribution of AE Aquarii. {\it Mon. Not. R. Astron. Soc.\/} {\bf 421}, 1557-1586 (2012)
\item S. B. Potter et al., Polarized QPOs from the INTEGRAL polar IGRJ14536-5522 (=Swift J1453.4-5524). {\it Mon. Not. R. Astron. Soc.\/} {\bf 402}, 1161-1170 (2010)
\item A. S{\l}owikowski, G. Kanbach, M. Kramer, A. Stefanescu, Optical polarization of the Crab pulsar: precision measurements and comparison to the radio emission.  {\it Mon. Not. R. Astron. Soc.\/} {\bf 397}, 103-123 (2009).
\item A. Karastergiou, S. Johnston, Absolute polarization position angle profiles of southern pulsars at 1.4 and 3.1 GHz. {\it Mon. Not. R. Astron. Soc.\/}  {\bf 365}, 353-366 (2006).
\item V. Radhakrishnan, D.J. Cooke D. J., Magnetic poles and the polarization structure of pulsar radiation. {\it Astrophys. J. Lett.\/} {\bf 3}, 225-229 (1969).
\item M. Landolfi, E. Landi Degl'Innocenti, M. Landi Degl'Innocenti, J.L, Leroy, Linear polarimetry of Ap stars I. A simple canonical model.  {\it Astron. Astrophys.} {\bf 272}, 285-298 (1993).
\item O.W. Butters et al., Circular polarization survey of intermediate polars I. Northern targets in the range 17 h $<$ RA $<$ 23. {\it Astron. Astrophys.} {\bf 496}, 891-902 (2009)
\item S.B Potter et al., On the spin modulated circular polarization from the intermediate polars NY Lup and IGR J15094–6649. {\it Mon. Not. R. Astron. Soc.\/} {\bf 420}, 2596-2602 (2012)
\item D. de B\'urca, A. Shearer, Circular polarization of synchrotron radiation in high magnetic fields.  {\it Mon. Not. R. Astron. Soc.} {\bf 450}, 533-540 (2015).
\item Ferrario, L., de Martino, D., G\"ansicke, B.T., Magnetic White Dwarfs,  {\it Space Sci Rev} {\bf 191}, 111-169 {doi: 10.1007/s11214-015-0152-0}, (2015).
\item J.I. Katz, AR Sco: A white dwarf synchronar, {\it arXiv: 1609.07172v1[astro-ph.SR], 22 Sept 2016} 
\item P.J. Meintjes, E. Jurua, Secondary star magnetic fields in close binaries. {\it Mon. Not. R. Astron. Soc.} {\bf 372}, 1279-1288 (2006)
\item W. Becker, J. Tr\"{u}mpher, The X-ray luminosity of rotation-powered pulsars, {\it Astron. Astrophys.} {\bf 326}, 682-691 (1997)
%\item R.N. Manchester, J.H. Taylor, Pulsars. (W.H. Freeman, San Francisco) (1977)
\item W. Becker, X-ray emission from pulsars and neutron stars, {\it Astrophysics and Space Science Library: Neutron Stars and Pulsars} {\bf 357} (Springer-Verlag Berlin Heidelberg) (ed. W Becker) pp. 91-140 (2009)
\item A.A. Abdo et al.,  The first Fermi Large Area Telescope catalog of gamma-ray pulsars, {\it Astrophys. J. Suppl. S.} , {\bf 187}, 460-494 (2010)

\item J-J Geng, B Zhang \& Y-F Huang, A model of white dwarf pulsar AR Scorpii, 
{\it Astrophys. J. Lett.\/} {\bf 831}, 10-14 (2016)
\item J.-F. Donati, J.D. Landstreet, Magnetic fields of nondegenerate stars. {\it Ann. Rev. Astron. Astrophys.} {\bf 47}, 333-370 (2009)
\item L. Mestel, Magnetic braking by a stellar wind I. {\it Mon. Not. R. Astron. Soc.}, {\bf 138}, 359-391 (1968)
\item L. Mestel, H.C. Spruit, On magnetic braking of late-type stars. {\it Mon. Not. R. Astron. Soc.}, {\bf 226}, 57-66 (1987) 
\item G. Haerendel, Acceleration from field-aligned potential drops, {\it Astrophys. J. Suppl. S.} {\bf 90}, 765-774 (1994).
\item L.A. Venter, P.J. Meintjes, The tenuous x-ray corona enveloping AE Aquarii, {\it Mon. Not. R. Astron. Soc.} {\bf 378}, 681-690 (2007).
\item H. van der Laan, A model for variable extragalactic radio sources, {\it Nature\/} {\bf 211}, 1131-1133 (1966).
\item T.S. Bastian, G.A. Dulk, G. Chanmugam, Radio flares from AE Aquarii: A low-power analog to Cygnus X-3?, {\it Astrophys. J.} {\bf 324}, 431-440 (1988).
\item J. Arons, E.T. Scharlemann, Pair formation above polar caps - structure of the low altitude acceleration zone. {\it Astrophys. J.} {\bf 231}, 854-879 (1979)
\item F.C. Michel, Coherent neutral sheet radiation from pulsars, {\it Comments Astrophys. Space Phys.} {\bf 3} 80-86 (1971)
\item F.V. Coroniti, Magnetically striped relativistic magnetohydrodynamic winds: The Crab nebula revisited {\it Astrophys. J.} {\bf 349} 538-545 (1990)
\item J.G. Kirk, Y. Lyubarsky \& J. P\'{e}tri, The Theory of Pulsar Winds and Nebulae, {\it Astrophysics and Space Science Library: Neutron Stars and Pulsars} {\bf 357} (Springer-Verlag Berlin Heidelberg) (ed. W Becker) pp. 421-450 (2009)
\item J. Patterson, The DQ Herculis Stars, {\it Pubbl. Astron. Soc. Pac.} {\bf 106}, 209-238 (1994)
\item P.J. Meintjes, On the evolution of the nova-like variable AE Aquarii, {\it Mon. Not. R. Astron. Soc.} {\bf 336}, 265-275 (2002).
\item K. Schenker, A.R. King, U. Kolb, G.A. Wynn, Z. Zhang, AE Aquarii: How cataclysmic variables descend from supersoft binaries,  {\it Mon. Not. R. Astron. Soc.} {\bf 337}, 1105-1112 (2002)

\end{enumerate}
\end{quote}

\begin{figure}

	\begin{center}
%	  \centering
	   \includegraphics[width=0.75\textwidth]{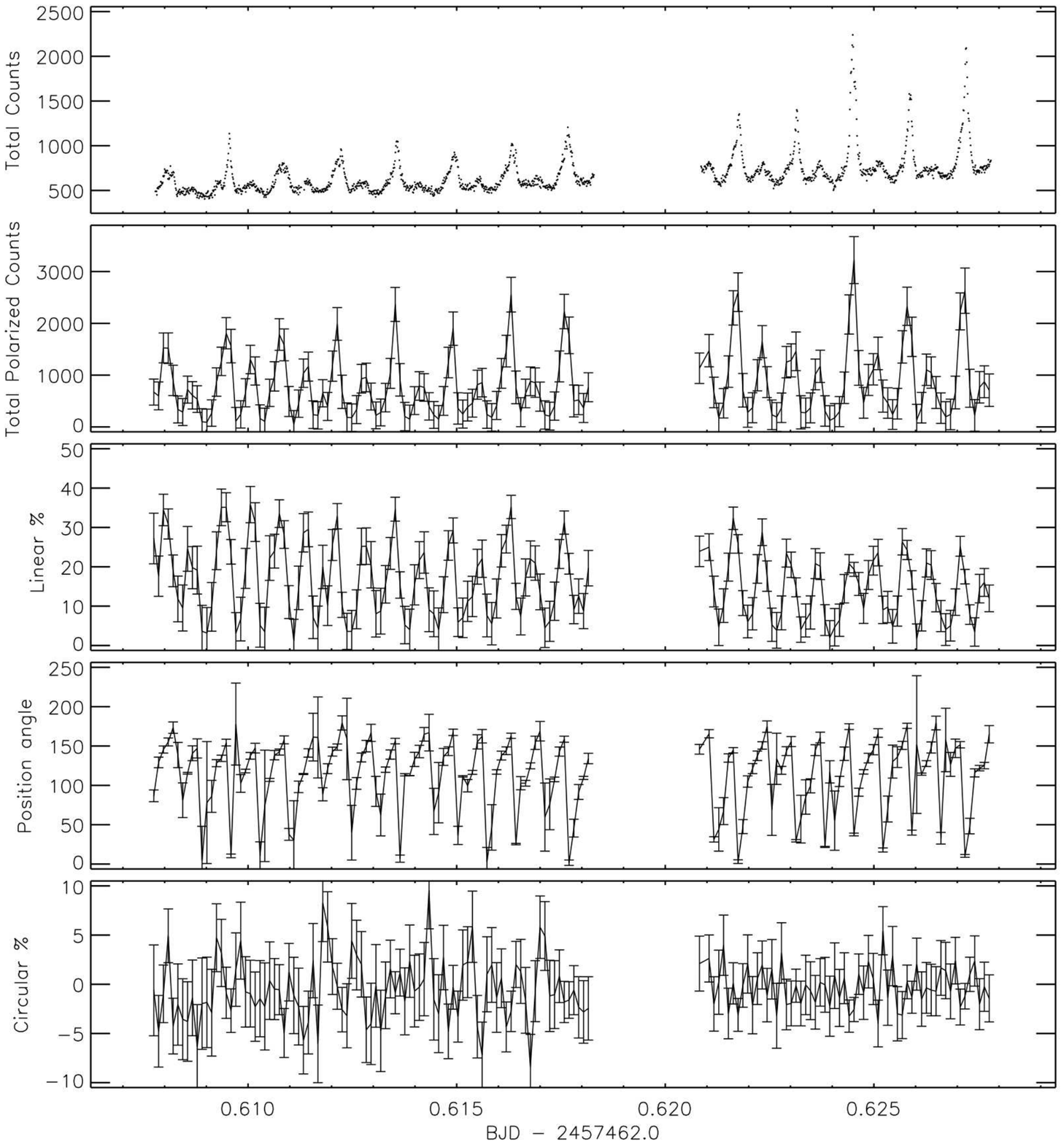}
\captionsetup{skip=-50pt}
		\caption{Time series polarimetry. \newline Red band (570 -- 900 nm) photopolarimetry of AR Sco taken on 14 March 2016. The top panel shows  the total intensity,  in 1 s bins, while the remaining three panels show, respectively, total polarized flux ({\it s}), degree of linear polarization ({\it p}), position angle of linear polarization ({\it $\theta$}) and the degree of circular polarization ({\it V/I}), all with 10 s bins. Counts in the top two panels are per time bin and error bars are 1$\sigma$. The data cover orbital phase interval $\phi = 0.10-0.23$ and the gap is when a background measurement was obtained. }
		\label{Fig. 1}
	\end{center}
\end{figure}

\begin{figure}

	\begin{center}
%	  \centering
	   \includegraphics[width=0.8\textwidth]{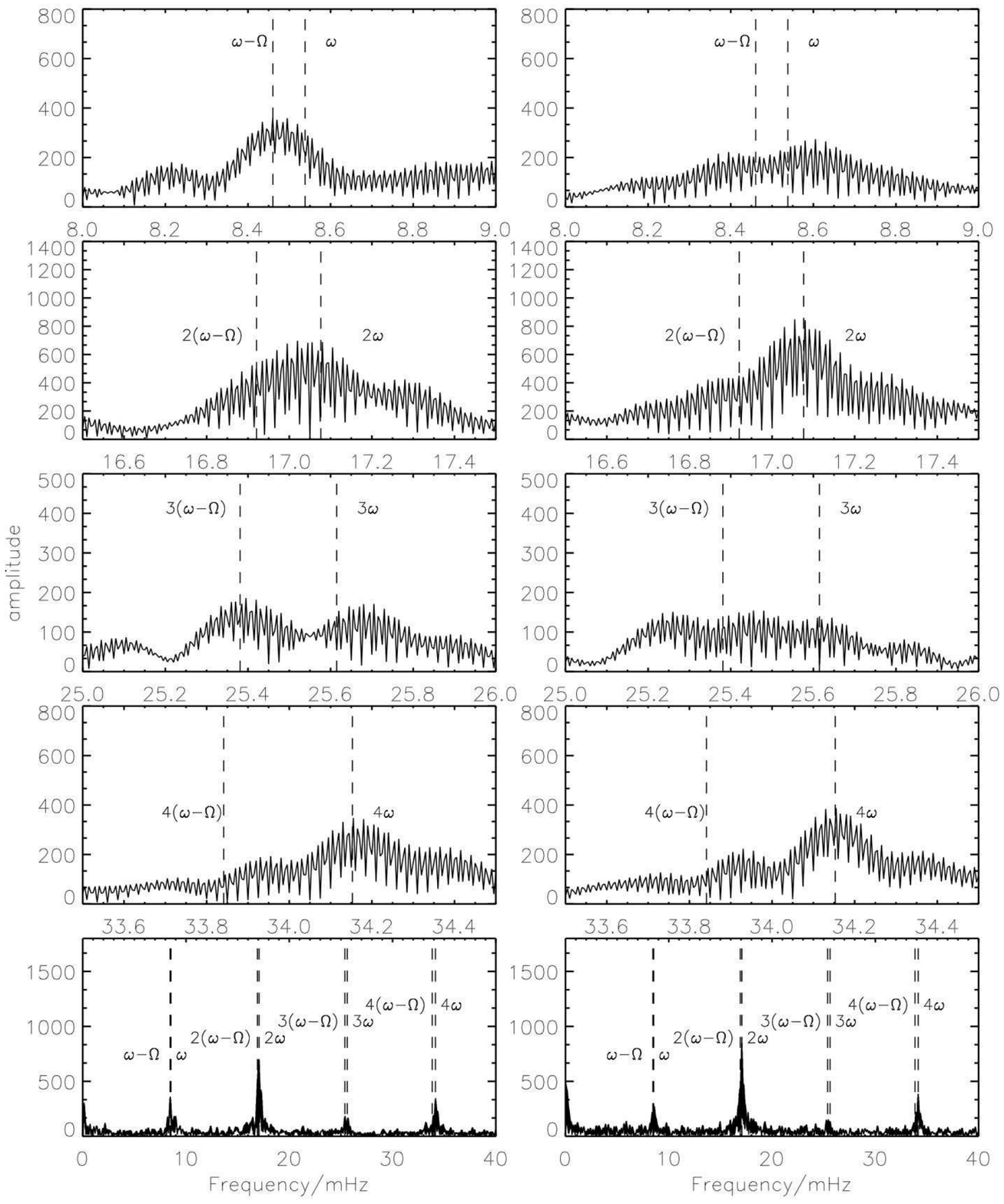}
\captionsetup{skip=-30pt}
		\caption{Polarimetry periodograms. \newline Amplitude spectra of the Stokes {\it U} (left) and {\it Q} (right) parameters from the two combined nights of HIPPO data, through the ``clear" filter (350 -- 900 nm). The bottom two panels cover frequencies from $0 - 40$ mHz (periods $<25$ s), while the top eight panels show details of the fundamental and harmonic frequencies in more detail. The dashed vertical lines mark the beat and spin frequencies ($\omega - \Omega$  and $\omega$, respectively) and their first, second and third harmonics.}
		\label{Fig. 2}
	\end{center}
\end{figure}

\begin{figure}
	\begin{center}
%	  \centering
	   \includegraphics[width=0.85\textwidth]{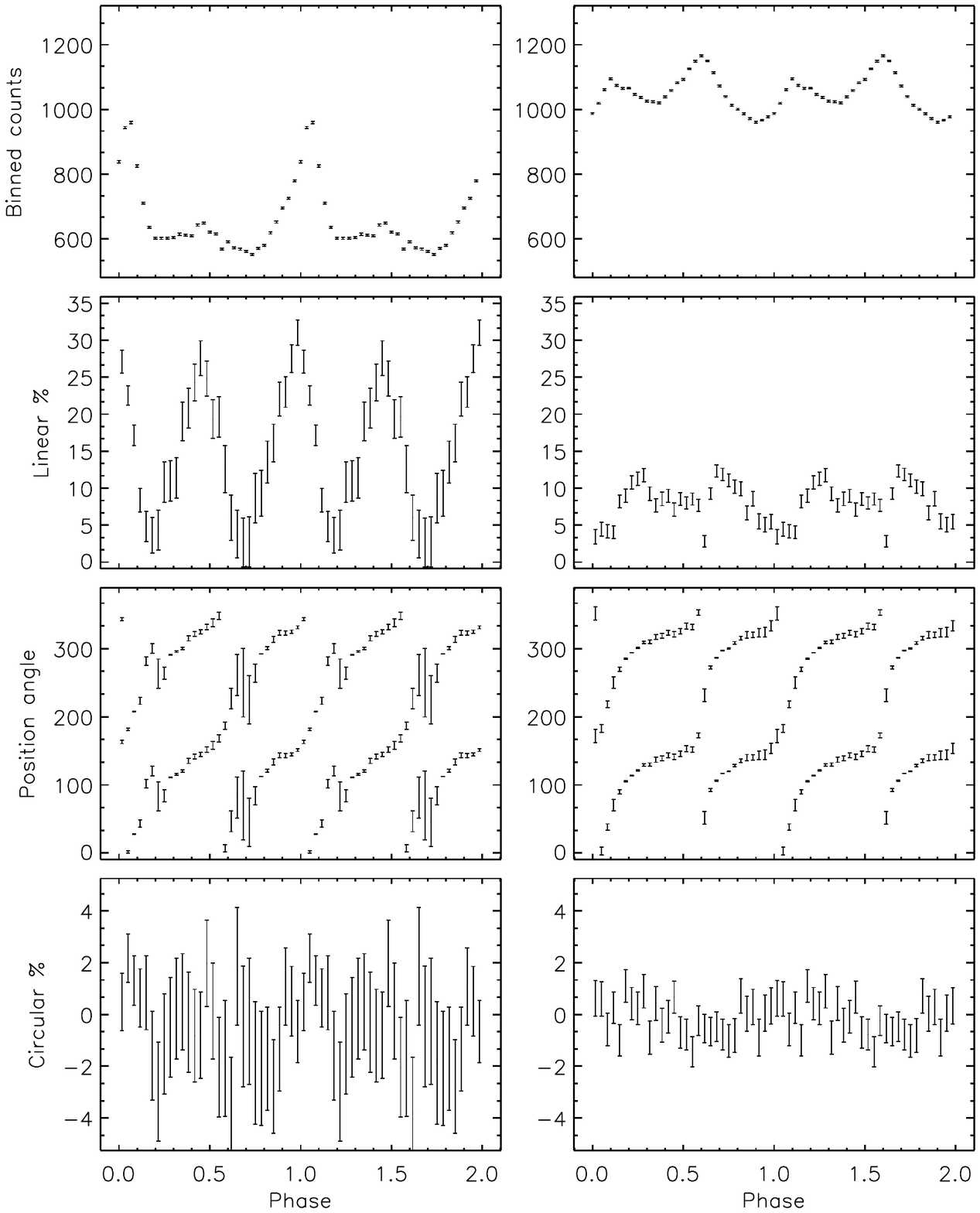}
\captionsetup{skip=-10pt}
		\caption{Spin modulated polarization. \newline The left and right panels show the spin phase folded (into 30 bins per cycle) red-band (OG570) variations for total intensity ({\it I}), degree of linear polarization ({\it p}), position angle ($\theta$)  and degree of circular polarization {(\it v}) on the 14/15 (left panels) and 15/16 March 2016 (right panels), respectively. Two cycles are shown for clarity. Error bars are 1$\sigma$.}
		\label{Fig. 3}
	\end{center}
\end{figure}

\begin{figure}
	\begin{center}
%	  \centering
	   \includegraphics[width=1.05\textwidth]{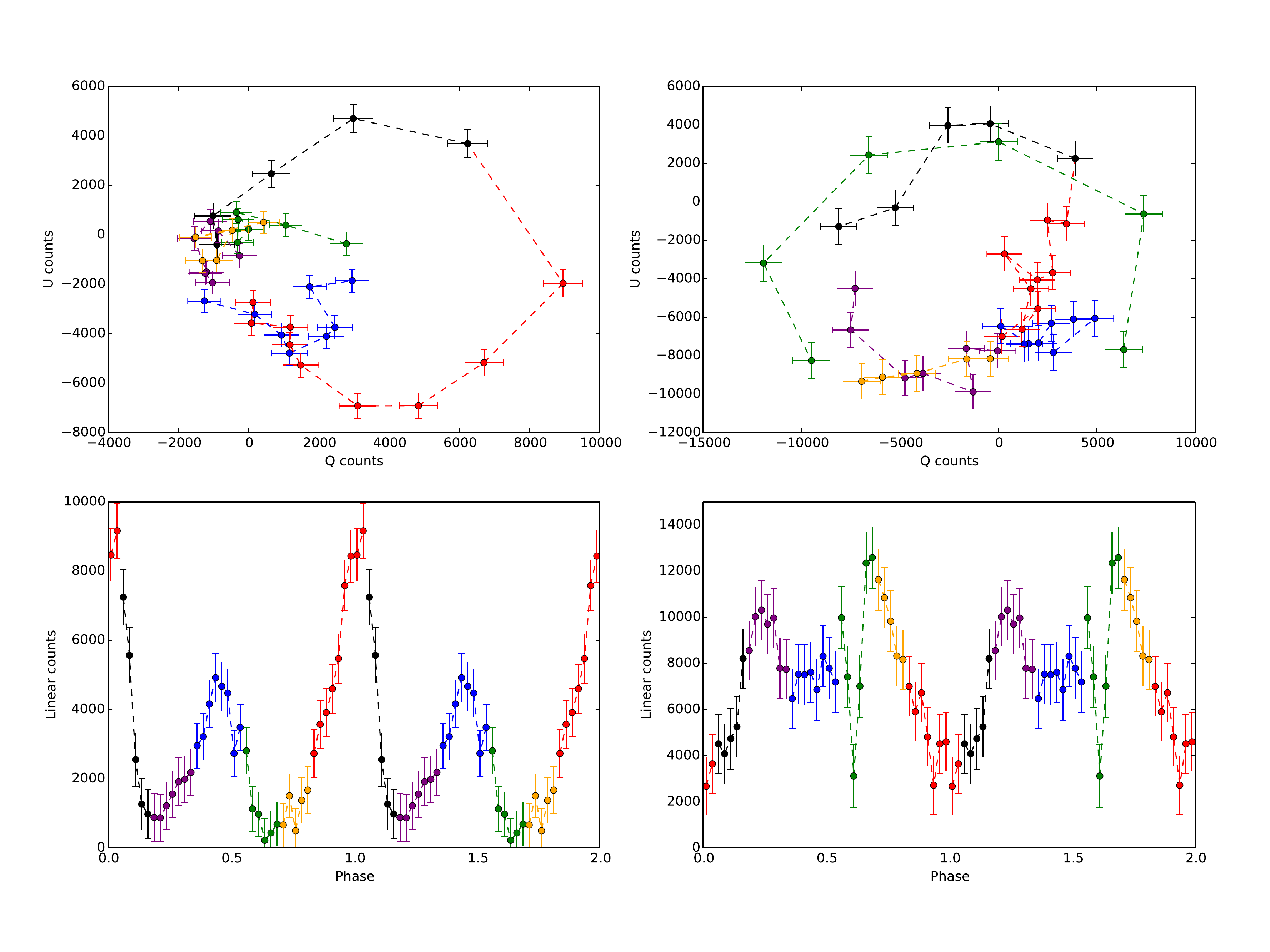}
\captionsetup{skip=0pt}
		\caption{{\it Q, U} and polarized flux variations over spin period. \newline 
Spin phased Stokes {\it Q} and {\it U} parameters (upper panels) and total linearly polarized flux (lower panels) for the red-band (570$-$900nm) on the two nights 14 and 15 March 2016. The migration of the Stokes {\it Q} and {\it U} amplitude pairs are shown, plotted every 3 s, and 40 points plotted per spin cycle. Points are colour-coded as in the average phase-folded linearly polarized flux plots.  {\it Q , U} pairs follow counter-clockwise trajectories.}

		\label{Fig. 4}
	\end{center}
\end{figure}

\newpage
\begin{center}
\section*{Methods}
\end{center}
\noindent
{\bf Data Sources:}
\newline  The data presented in this paper arise from observations taken at the South African Astronomical Observatory using the 1.9 m telescope.  High speed photopolarimetry of AR Sco was conducted utilizing the HIPPO photopolarimeter on two consecutive nights, on 14 and 15 March 2016 (see Supplementary Table 1 for observing log). HIPPO is a two-channel all-Stokes photon counting photopolarimeter ({\it 7\/}) which records all observations as intensity measurements every millisecond. Rotating half and quarter waveplates modulate the signal at 10 Hz, and the data are accumulated in 100 bins per waveplate rotation. The Stokes parameters are then determined by a Fourier decomposition of the modulated signal; the $4\theta$ and $8\theta$ terms determining {\it Q} and {\it U} and the 6$\theta$ term determining {\it V} (N.B. $\theta$ here should not be confused with the position angle of linear polarization).  The total intensity (Stokes {\it I}) data were post-processed by summing the millisecond measurements to a time resolution of 1 s. Likewise, the 100 millisecond polarization arrays were summed to a time resolution of 10 s. Calibration observations of the polarized standard star HD160529 ({\it 36\/}) were observed on both nights to determine the zero points for both waveplate angles. In addition, an observation was made of a nearby star, $\sim$40 arcsec north east of AR Sco in order to estimate the level of background polarization.

Simultaneous observations were undertaken in two broad wavebands defined by the convolution of the intrinsic response of the GaAs photomultipliers with the respective filters used. For both observations we used a clear fused silica filter (essentially no filter) and a Schott OG 570 glass filter, respectively, for the two channels. Channel 1 had a bandpass of 350$-$900 nm, while channel 2 covered 570$-$900 nm. All of the data were converted to Barycentric Julian Date using the accurate prescription presented in ({\it 37}). In  Supplementary Figures 1 and 2 we show the variation of the Stokes parameters {\it I, Q, U \& V} for the broad band filter on both nights, while in Supplementary Figures 3 to 6 we show the polarized flux ({\it s}), degree of linear polarization ({\it p}) and position angle ($\theta$) for both night and both filters.

\noindent
\newline
{\bf Period analysis:}
\newline  The reduced HIPPO photopolarimetry was analysed using a discrete Fourier transform (DFT) periodogram program.  The two consecutive nights of data (14 and 15 March) were combined and power spectra derived for i) the Stokes {\it Q} and {\it U} parameters (Figure 2), ii) the total and polarized intensity, iii) the degree of linear polarization {\it (p)} and the position angle of the linear polarization, i.e. the E-vector ($\theta$).  The periodograms suffer from significant aliasing due to the relatively short spin (118.2 s) and beat (117.1 s) periods, which are unresolved on single nights, and because of the $\sim$24 h data gap, results in significant cycle count ambiguity. 

Just as was seen in the original study ({\it 2\/}) for the $g^\prime$ band variability, the total intensity ({\it I}) varies predominantly at the beat frequency and its first harmonic (Supplementary Figure 7, right panels), with the latter being the stronger of the two. In addition, the harmonic of the spin frequency is also clearly seen, but at a lower power compared to the beat harmonic.  The power spectra of the UV light curves derived from the Hubble Space Telescope observations ({\it 2\/}), although strongly aliased, clearly show that the power in the first harmonics of the spin and beat frequencies are approximately equal, implying that the white dwarf’s magnetic field is directly (through beaming) or indirectly (through reprocessing) modulating the UV emission. In the case for the polarized flux (Supplementary Figure 7, left panels), we see a broad spread of power at the first harmonic, consistent with power at both spin and beat harmonic frequencies. In the case of the 3rd harmonics, the spin component (4$\theta$) dominates, at 29.3 s. The reason that the power of the spin modulated polarized flux and {\it p} is less than for the beat modulation is attributed to the modulation of {\it I} occurring at two closely spaced periods (beat and spin), while {\it Q} and {\it U} vary predominantly at the spin. Thus any calculation involving {\it Q, U} and {\it I} (as for {\it p}), will introduce some additional power at the beat period.  

\noindent
\newline
{\bf Binary parameters:}
\newline 
AR Sco consists of a magnetic white dwarf and a M5 dwarf star orbiting one another with an orbital period of $P_{\rm orb} = 3.56 \, \mbox{h}$. The pulsations observed from the UV to the radio show a double peaked pulse profile, predominantly at the beat period $P_{\rm b}$, implying that the secondary has to be a site for spin modulated reprocessing. Furthermore, narrow line emission is seen which clearly follows the orbital motion of the secondary star and is situated near the L1 point. Based on the limits placed, the white dwarf has a mass of $M_{1} \approx \, 0.8 \, M_{\odot}$, while the secondary star has a mass of about $M_{2} \approx \, 0.3 \, M_{\odot}$ and therefore a mass ratio of $q = (M_{2}/M_{1}) = 0.375$. Sinusoidal variations in the optical/IR light suggest that the secondary star is being irradiated, with its hotter/brighter side facing the white dwarf, which is best seen at orbital phase $\phi_{orb} = 0.5$ (superior conjunction of the secondary). 

Assuming that the secondary is close to filling its Roche lobe, then the standard relationship between the Roche lobe size of the secondary ($R_{\rm L,2}$) and binary separation ($a$), applicable to binaries in the range $0.3 < q < 20$  ({\it 38\/}) is given by $R_{L, 2}/a = 0.38 + 0.2 \log q$. For AR Sco ($q = 0.375$) the orbital separation $a \sim 8 \times 10^{10} (M_{\rm wd}/0.8 \, M_{\odot})^{1/3} (P_{\rm orb}/3.56 \, \mbox{hr})^{2/3} \, \mbox{cm}$. Thus the Roche lobe radius of the secondary star is estimated to be $R_{L,2}  \sim 0.3 a \sim 2.53 \times 10^{10} \, \mbox{cm}$. The distance of the L1 region from the white dwarf primary can be estimated from the relation $(b1/a) = 0.5 - 0.227 \log q$ (for $0.1 < q < 10$) (see e.g. {\it 39}), resulting in $b1 \sim 0.6 a \sim 5 \times 10^{10} \, \mbox{cm}$. In comparison, the radius of the light cylinder, where the white dwarf magnetosphere co-rotates with the speed of light ($c= \omega_{\rm wd} \, r_{\rm l-c})$, is of the order of $r_{\rm l-c} \sim 6 \times 10^{11} \, (\omega_{\rm wd}/0.054 \, \mbox{rad s}^{-1})  \, \mbox{cm}$, implying that the binary system is inside the white dwarf pulsar's light cylinder. 

\noindent
\newline
{\bf Luminosities:}
\newline
It has been found ({\it 2}) that the white dwarf in AR Sco is currently spinning down on a timescale of $\tau_{\rm s-d} = (\nu_{\rm s}/\dot{\nu_{\rm s}}) \sim 10^{7} \, \mbox{yr}$. The spin-down luminosity, assuming a $0.8 M_{\odot}$ white dwarf, is of the order of $L_{\dot{\nu_{\rm s}}} = - 4 \pi^{2} I \nu_{\rm s} \dot{\nu_{\rm s}} = 1.5 \times 10^{33} \, \mbox{erg s}^{-1}$. The X-ray luminosity  is $L_{\rm x} = 4.9 \times 10^{30} \, \mbox{erg s}^{-1}$, implying $\alpha = (L_{\rm x}/L_{\dot{\nu_{\rm s}}}) = 3.27 \times 10^{-3}$. This is very similar to the ratio found in spin-powered pulsars as well as the white dwarf in the cataclysmic variable system AE Aquarii (Supplementary Figure 8), which has a ratio $\alpha \sim 10^{-3}$,  where $L_{\rm x} \sim 5 \times 10^{30} \, \mbox{erg s}^{-1}$ and $L_{\dot{\nu_{\rm s}}} = 6 \times 10^{33} \, \mbox{erg s}^{-1}$ (e.g. see references {\it 4, 5, 26}).

\noindent
\newline
{\bf Magnetic dipole radiation:}
\newline
The luminosity coming from magnetic dipole radiation (also known as Poynting flux) in AR Sco is given as:
\begin{eqnarray}
L_{\rm m-d} & = & \frac{2 B_{1,*}^{2} \omega_{\rm wd}^{4} R_{\rm wd}^{6} \sin^{2} \chi}{3 c^{3}},
\end{eqnarray}
where $B_{1,*}$, $\omega_{\rm wd}$, $R_{\rm wd}$ and $\chi$ represent the surface polar field of the white dwarf, the white dwarf spin angular velocity and radius, as well as the angle between the spin and magnetic axes. For our calculations of luminosities we used the Hamada-Salpeter relation to determine that the white dwarf has a radius of $R_{\rm wd} \sim 5.5 \times 10^{8} \, (M_{\rm wd}/0.8 \, M_{\odot})^{-0.8} \, \mbox{cm}$.

Re-arranging this equation to determine the magnetic field strength gives:
\begin{eqnarray}
B_{1, *} \, \sin\chi & = & \left(\frac{3 c^{3} L_{\dot{\nu_{\rm s}}} P_{\rm wd}^{4}}{2 (2 \pi)^{4} R^{6}_{\rm wd}}\right)^{1/2}.
\end{eqnarray}

\noindent
\newline
{\bf Spin-down torque:}
\newline
It has been shown (e.g. {\it 40, 41, 42\/}) that the rate of dissipation of magnetic energy in the surface layers of a cataclysmic variable secondary star of depth $\delta$ (skin-depth) is given by $\dot{W} = (B^{2}/8 \pi) (4 \pi R_{2}^{2} \delta)~ \omega_{\rm b}$, where $B$ is the estimated magnetic field of the white dwarf at the distance of the secondary star, $\omega_{\rm b}$ is the beat period angular frequency of the white dwarf (essentially the same as the spin angular frequency in AR Sco) and  $\delta = \sqrt{2 \eta_{\rm tur}/\omega_{\rm b}}$ represents the dissipation depth of magnetic energy, where $\eta_{\rm tur}$ is the turbulent plasma resistivity. 

The total magnetic torque exerted on the secondary star, i.e. $T_{\rm D} = (\dot W/\omega_{\rm b})$, results in a synchronization timescale which is given by $t_{\rm syn} = I \omega_{\rm b}/T_{\rm D}$, where {\it I} is the moment of inertia of the white dwarf, which is of the order of $I \sim 1.5 \times 10^{50} \, \mbox{g cm}^{2}$.  If we assume a 500 MG dipolar magnetic field for the white dwarf, the upper limit derived from the dipole radiation, the field strength at the distance of the secondary star is $B_{\rm b1} \sim 160 \, \mbox{Gauss}$. For $\delta \sim 10^{8} \, \mbox{cm}$, i.e. using  $\eta_{\rm tur} \leq 10^{15} \, \mbox{cm}^{2}\mbox{s}^{-1}$ (see e.g. {\it 17}), the magnetic torque exerted on the secondary star is of the order
$T_{\rm D}  = (B_{1,*}^{2}/8 \pi) (R_{\rm wd}/a)^{6} (4 \pi R_{2}^{2} ~\delta) \sim  10^{33} \, \mbox{erg}~ (= \, \mbox{dyn cm})$. The resulting timescale for spin-synchronization through MHD torques for a 500 MG white dwarf is therefore $t_{\rm syn} \sim 2.5 \times 10^{8} \, \mbox{yr}$. This supports the notion that the magnetic dissipation in the thin surface layers of the secondary star ($\delta \sim 0.004 R_{\rm L,2}$) will contribute to the dissipation of rotational  kinetic energy of the white dwarf, together with other mechanisms like magnetic dipole radiation. This will eventually lead to the sychronization of the two stars during the current low mass accretion phase. When accretion eventually switches on, once the M-star re-attaches to the Roche lobe, then AR Sco will become a synchronised magnetic cataclysmic variable, or polar,  consisting of a highly magnetized white dwarf rotating synchronously with its M5 secondary companion.

\noindent
\newline
\newline
{\bf Magnetic field interactions:}
\newline
It has been demonstrated (e.g. {\it 17}) that mass transferring secondary stars with orbital periods between $P_{\rm orb} = 3 - 4$ h can have surface polar magnetic fields of the order of $B_{\circ} \sim 3000 \, \mbox{G}$, which implies that the M5 secondary star of AR Sco will most probably be magnetically active. This is supported by MHD modelling in convective envelopes of dwarf stars (e.g. {\it 22}). This provides a mechanism for explaining the observed spectral energy distribution of AR Sco below $\nu \sim 10^{13}\, \mbox{Hz}$ in terms of magnetic energy dissipation in the field of the secondary star. 

The field-aligned electric potentials in the M-star magnetic flux tubes are of the order of
\begin{eqnarray}
\Phi_{||} & = & \left(\frac{300 (\Delta B_{\perp})^{3}}{64 \pi e n_{\rm e}^{3/2} \sqrt{4 \pi k T}}\right) \, \mbox{V} 
\nonumber \\
\\
& \sim & 200 \, \left(\frac{\Delta B_{\perp}}{700 \, \mbox{G}}\right)^{3} \left(\frac{n_{\rm e}}{10^{10} \, \mbox{cm}^{-3}}\right)^{-3/2} \left(\frac{T}{10^{4} \, \mbox{K}}\right)^{-1/2} \, \mbox{MV},
\end{eqnarray}
where we assumed $\Delta B_{\perp} \rightarrow B_{\rm b1} \sim 700 \, \mbox{G}$ represents the perturbation induced in the coronal loops, or flux tubes, of the secondary as a result of the white dwarf field periodically pushing against it. These coronal structures will be periodically perturbed, pushed sideways by the rotating white dwarf field and generating a ``fracture zone" which will induce large field-aligned potentials that can accelerate charged particles (see {\it 26\/}). 

In these interactions, the magnetic pitch can increase $\gamma_{\rm \phi} = (\Delta B_{\perp}/B) \rightarrow 10$ (see e.g. {\it 43, 44\/}) before the fields become unstable to reconnection. It should be mentioned that the magnetic field perturbation is still below the maximum magnetic field near the poles of the secondary, which can be of the order of 3000 G (see e.g. {\it 17}). It can be shown that electrons with energies of the order of $\gamma_{\rm e} \leq 400$ trapped in pumped coronal fields of the order of a $\langle B_{\rm cor} \rangle \sim  100 \, \mbox{G}$ (at a distance from the secondary surface of $\sim 2 \times R_2$) will radiate synchrotron radiation at frequencies up to $\nu_{\rm syn} \sim \gamma^{2}_{\rm e} e B/(2 \pi m_{\rm e} c) \leq 4 \times 10^{13} (\gamma_{\rm e}/400)^{2} (B/100 \, \mbox{G}) \, \mbox{Hz}$, i.e. in the infrared to radio regime.

\noindent
\newline
{\bf Pulsar-like particle acceleration:}
\newline
For AR Sco, electrical potentials of the order
\begin{eqnarray}
\Delta V & \sim & 10^{12} \left(\frac{P_{\rm wd}}{117 \,  \mbox{s}}\right)^{-5/2} \left(\frac{\mu}{8 \times 10^{34} \, \mbox{G cm}^{3}}\right) \left(\frac{R_{*}}{5.5 \times 10^{8} \, \mbox{cm}}\right) \, \mbox{V}
\end{eqnarray}     
can be induced between the white dwarf and the light cylinder ({\it 29}), i.e. where the white dwarf magnetosphere co-rotates at the speed of light ($c= \omega_{\rm wd} \, r_{\rm l-c})$. For AR Sco the light cylinder radius is of the order of $r_{\rm l-c} \sim 6 \times 10^{11} \, (\omega_{\rm wd}/0.054 \, \mbox{rad s}^{-1})  \, \mbox{cm}$. Since this electric field is several orders of magnitude stronger than gravity, charged particles like electrons can be pulled from the surface of the white dwarf and accelerated to relativistic energies ({\it 45}), of the order of $\gamma_{\rm e} \sim 10^{6}$, resulting in synchrotron emission up to frequencies of the order of 

\begin{eqnarray}
 \nu_{\rm syn} & \sim & 10^{18} \left(\frac{B_{\rm l-c}}{0.4 \, \mbox{G}}\right) \left(\frac{\gamma_{\rm e}}{10^{6}}\right)^{2} \, \mbox{Hz},
\end{eqnarray}

with peak emission occurring  $\nu_{\rm c} \sim 0.3 \nu_{\rm syn} \sim 3 \times 10^{17} \, \mbox{Hz}$, in the X-ray regime. 

Additionally, outside the light cylinder the fast rotating magnetosphere will turn into a relativistic magnetohydrodynamic ``striped" wind ({\it 30, 31, 32} and references therein). It will consist of zones of opposite magnetic polarity that will form current sheets, which can be a source of particle acceleration through magnetic reconnection and pulsed incoherent synchrotron emission (see {\it 32} for a review). The emission will be pulsed as a result of a relativistic beaming effect due to these zones of opposite magnetic polarity that propagate outward at relativistic velocities ($v \rightarrow c$), separated by $\Delta l = \pi r_{L}$ (here $r_{\rm L}$ is the light cylinder radius). Each outward propagating zone is beaming radiation into a forward cone of opening angle $\theta \sim 2 / \gamma$, where $\gamma$ represents the wind Lorentz factor. Pulses will be observed when the time delay between emission on each expanding wave front that intersects the observer, i.e. $\Delta t \sim R/2 \gamma^{2} c$), is less than the time delay between the radiation from two successive outward propagating wave fronts ($\Delta T = \Delta l/c$). 

A characteristic of the synchrotron radiation from the striped wind outside the light cylinder radius is the fact that the toroidal magnetic field component oscillates in the same sense as the rotating pulsar, which results in pulsations at both the fundamental and first harmonic of the spin period. The emission is also expected to be polarized, emitting both linear and a small component of circular polarized light ({\it 32}). Applied to the Crab pulsar, numerical simulations of the expected pulse profiles produced in a striped wind mimic exactly the observed  pulse profiles,  together with the position angle swings through  180$^{\circ}$. The observed polarized pulse profiles of AR Sco (Figures 3 and 4), especially near orbital phase $\Phi_{o}  \sim$ 0.15, mimic that of the Crab pulsar, perhaps implying that a striped pulsar wind is active in AR Sco. The fact that the striped wind occurs in a binary system may further introduce a beat frequency in the pulsations.

Both these processes, namely synchrotron emission in the vicinity of the light cylinder, or outside in the striped pulsar-like wind, can account for the nature of the polarized pulsed  emission seen in optical lightcurves of AR Sco. Both processes can explain the polarization, i.e. a signature of synchrotron radiation, pulsed at both the white dwarf spin period and the beat (synodic) period.

\noindent
\newline

\begin{quote}

\section*{Methods References}
\begin{enumerate}
\setcounter{enumi}{35}
\item J. Hsu J., M. Breger, On standard polarized stars. {\it Astrophys. J.} {\bf 262}, 732-738 (1982).
\item J. Eastman, R. Siverd, S. Gaudi, Achieving better than 1 minute accuracy in the heliocentric and barycentric Julian dates. {\it Publ. Astron. Soc. Pacific} {\bf 122}, 935-946 (2010).
\item B.Paczynski, Evolution of Single Stars. VI. Model Nuclei of Planetary Nebulae, {\it Acta Astronomica} {\bf 21}, 417-435 (1971a).
\item M. Plavec,  P. Kratochvil, Tables for the Roche model of close binaries, {\it Bull. Astr. Czech.} {\bf 15}, 165-170 (1964).
\item P.C. Joss , J.I. Katz, S.A. Rappaport, Synchronous rotation in magnetic x-ray binaries,  {\it Astrophys. J.\/} {\bf 230}, 176-183 (1979).
\item J. Papaloizou, J.E. Pringle, A model for VW Hydri. {\it Astron. Astrophys.} {\bf 70}, L65-67 (1978).
\item C.G. Campbell, {\it Magnetohydrodynamics in Binary Stars\/} (Kluwer Academic Publishers, Dordrecht), p. 88-89 (1997).
\item P Ghosh, F.K. Lamb, Plasma physics of accreting neutron stars, {\it Neutron Stars: Theory and Observation} (Kluwer Academic Publishers, Dordrecht) (eds. J.E. Ventura and D. Pines) , p. 363-444 (1991).
\item P.J. Meintjes, O.C. de Jager, Propeller spin-down and non-thermal emission from AE Aquarii, {\it Mon. Not. R. Astron. Soc.} {\bf 311}, 611-620 (2000).  
\item P. Goldreich, W.H. Julian, Pulsar Electrodynamics, {\it Astrophys. J.} {\bf 157} 869-880 (1969)

\end{enumerate}
\end{quote}

\newpage

\begin{center}
\section*{Supplementary Tables \& Figures:}
\end{center}
\vspace{120pt}

\renewcommand{\tablename}{Supplementary Table}

\begin{table}[!hp]
%\centering
\caption{AR Sco Photopolarimetry Observing Details}\label{tabmag}
\begin{tabular}{c|c|c|c|c|c|c}
\hline
\,
Telescope / & Observation & Date & HJD start$^*$ & HJD end$^*$ & Duration & Orbital \\
Instrument & Type & & & & (h) & Phase$^{\#}$\\
   \hline

SAAO 1.9-m	& All-Stokes &	2016-03-14 &	7462.604	& 7462.628 &	0.57	& 0.07$-$0.23\\
/ HIPPO		 &  & 2016-03-15 &	7463.541 &	7463.611	& 1.68	& 0.38$-$0.85\\

\hline
\end{tabular}
\newline
\hfill \break
$^*$ HJD - 2450000\\
$^{\#}$ from orbital ephemeris ({\it 2})

\end{table}

\renewcommand{\figurename}{Supplementary Figure}
\setcounter{figure}{0}

\begin{figure}
	\begin{center}
	   \includegraphics[width=1\textwidth]{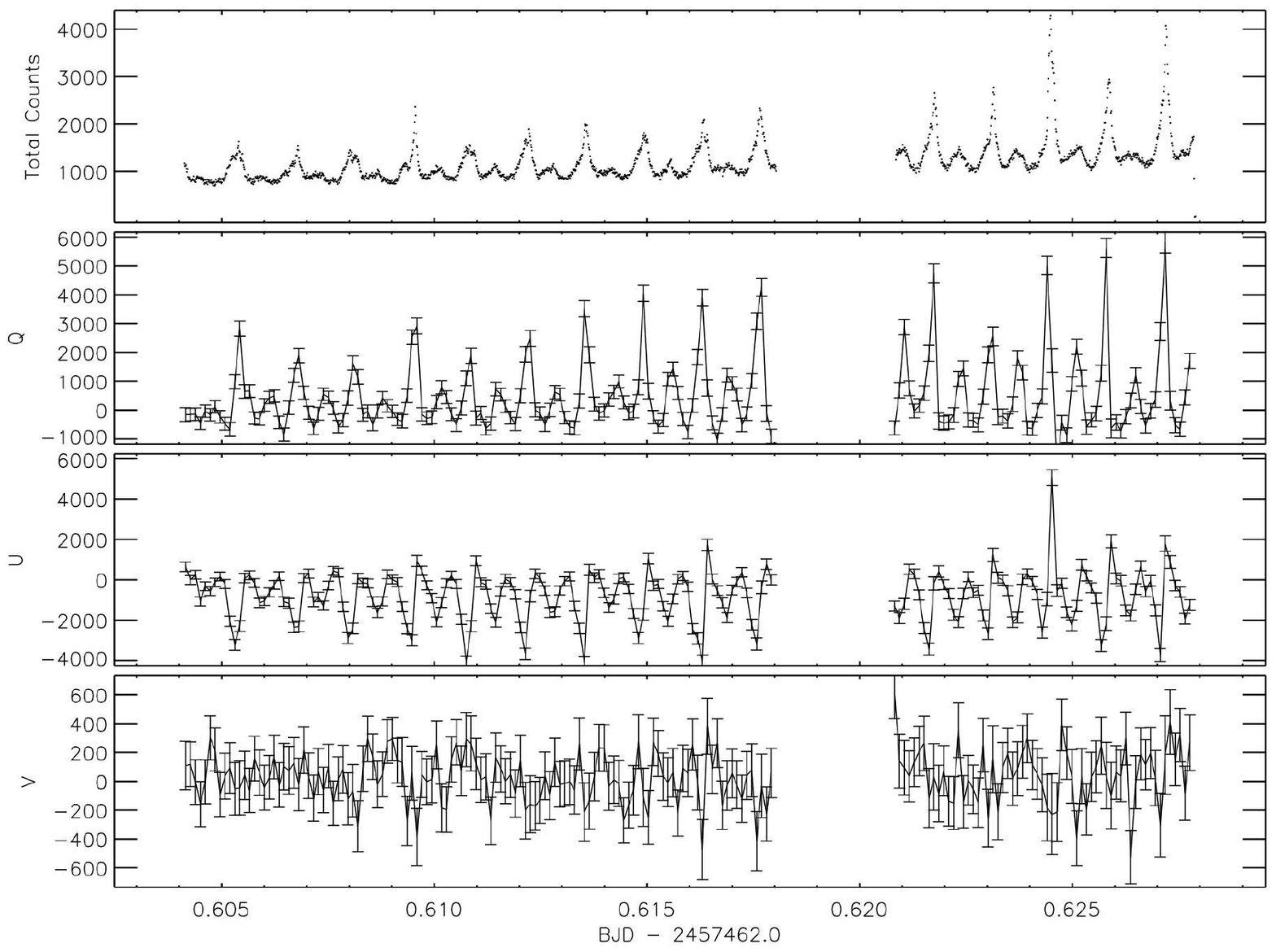}
\captionsetup{skip=-250pt}
		\caption{Time series polarimetry. \newline Broad band (340 -- 900 nm) photopolarimetry of AR Sco taken on 14 March 2016. The panel show, from the top,  the total intensity ({\it I}) in 1 s bins and the Stokes {\it Q, U \& V} values in 10 s bins, where the error bars are $\pm 1\sigma$.  The data cover orbital phase interval $\phi = 0.07-0.23$ and the gap is when a background sky measurement was obtained. }
		\label{Fig. S1}
	\end{center}
\end{figure}

\newpage
\begin{figure}
	\begin{center}
	   \includegraphics[width=1\textwidth]{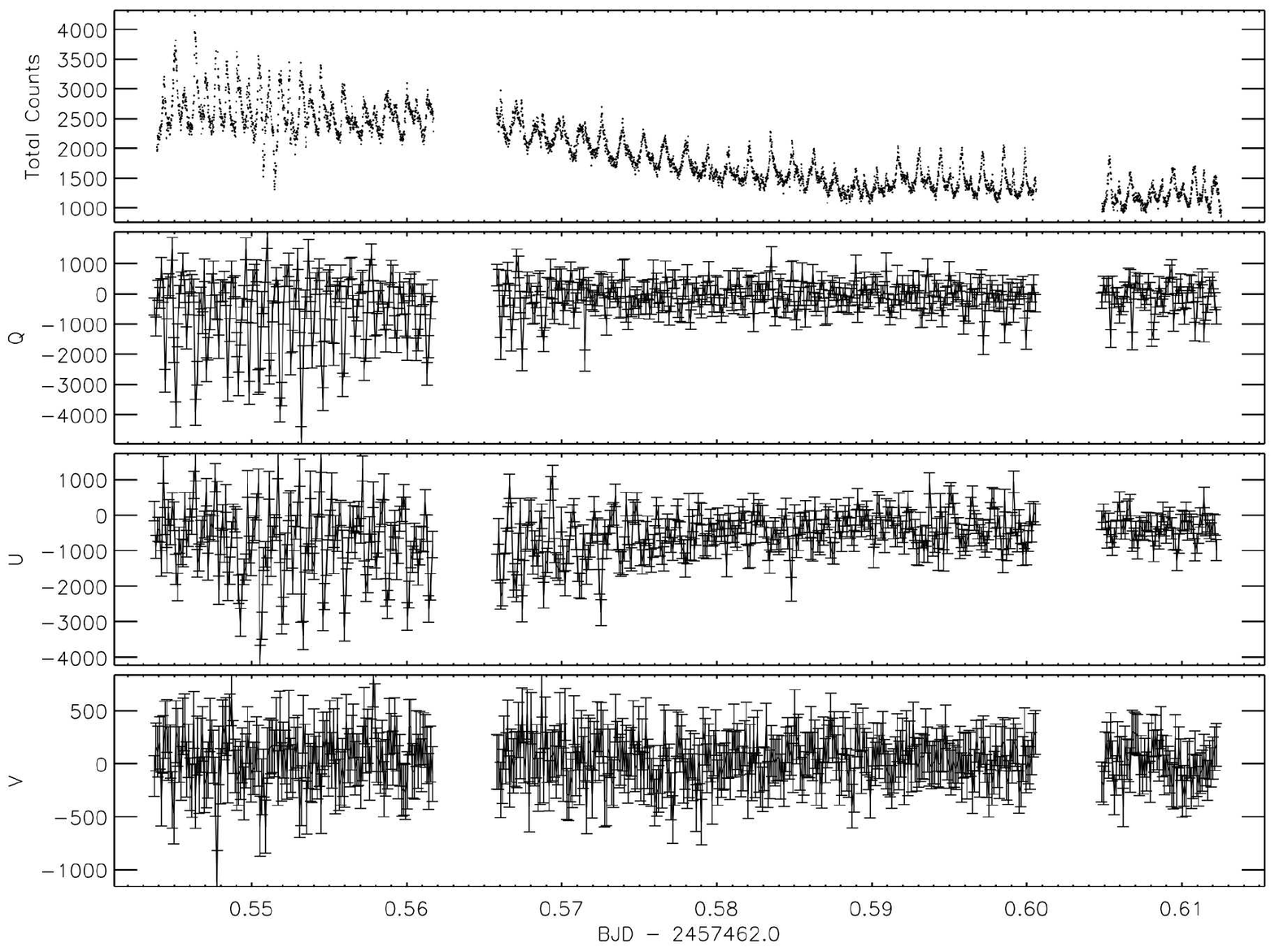}
\captionsetup{skip=-250pt}
		\caption{Time series polarimetry. \newline Broad band (340 -- 900 nm) photopolarimetry of AR Sco taken on 15 March 2016. The panel show, from the top,  the total intensity ({\it I}) in 1 s bins and the Stokes {\it Q, U \& V} values in 10 s bins, where the error bars are $\pm 1\sigma$.  The data cover orbital phase interval $\phi = 0.38-0.85$ and the gaps are when background sky measurements were obtained. }
		\label{Fig. S2}
	\end{center}
\end{figure}

\newpage
\begin{figure}
	\begin{center}
	   \includegraphics[width=0.85\textwidth]{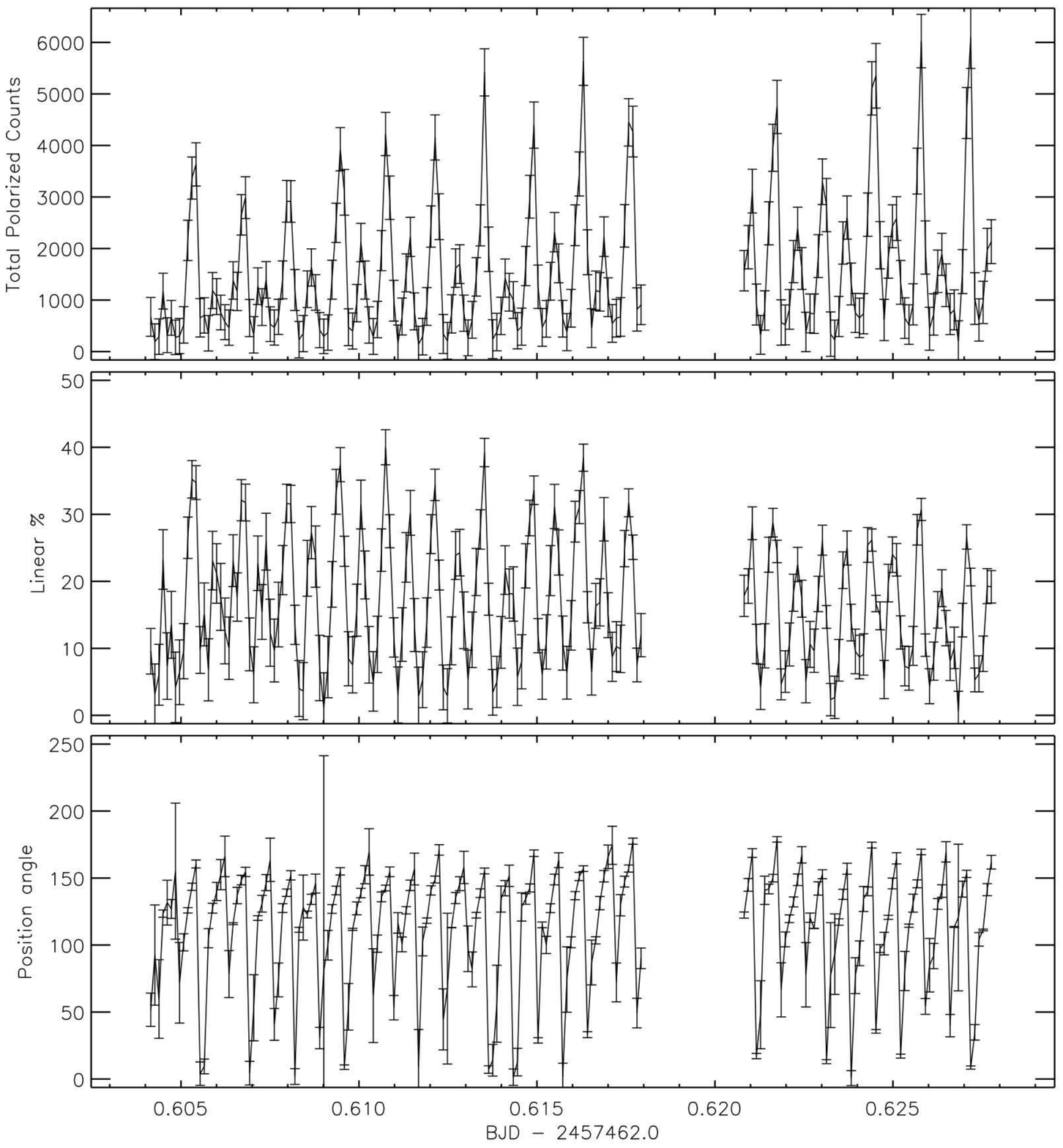}
\captionsetup{skip=-40pt}
		\caption{Time series polarimetry. \newline Broad band (340 -- 900 nm) photopolarimetry of AR Sco taken on 14 March 2016, in 10 s bins. The panel show, from the top,  the total polarized flux ({\it s}), degree of linear polarization ({\it p}) and position angle of linear polarization ({\it $\theta$}). The error bars are $\pm 1\sigma$.  The data cover orbital phase interval $\phi = 0.07-0.23$ and the gap is when a background sky measurement was obtained.  }
		\label{Fig. S3}
	\end{center}
\end{figure}

\newpage
\begin{figure}
	\begin{center}
	   \includegraphics[width=0.85\textwidth]{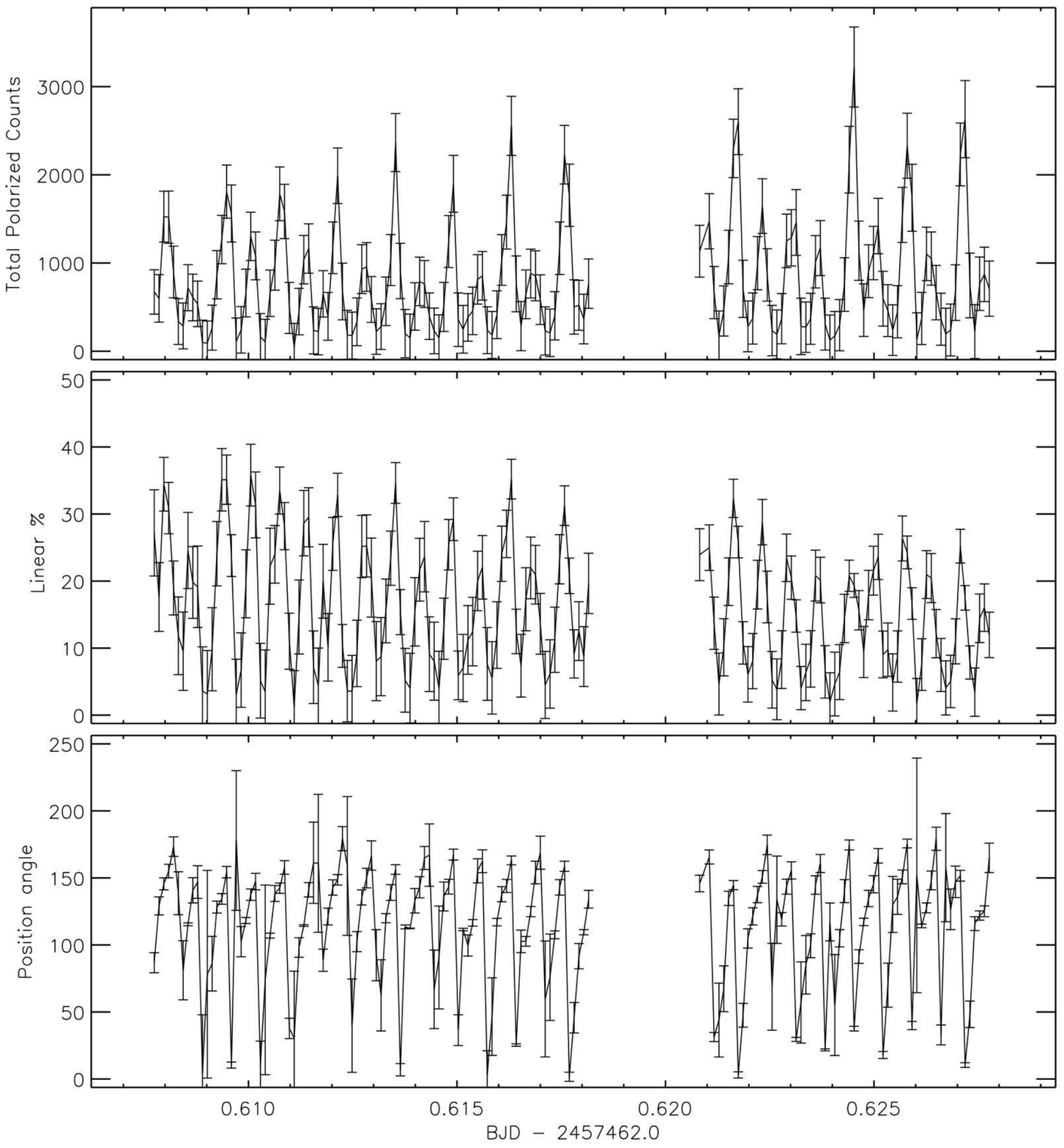}
\captionsetup{skip=-40pt}
		\caption{Time series polarimetry. \newline Red band (570 -- 900 nm) photopolarimetry of AR Sco taken on 14 March 2016, in 10 s bins. The panels show, from the top,  the total polarized flux ({\it s}), degree of linear polarization ({\it p}) and position angle of linear polarization ({\it $\theta$}). The error bars are $\pm 1\sigma$.  The data cover orbital phase interval $\phi = 0.10-0.23$ and the gap is when a background sky measurement was obtained. }
		\label{Fig. S4}
	\end{center}
\end{figure}

\newpage
\begin{figure}
	\begin{center}
	   \includegraphics[width=0.85\textwidth]{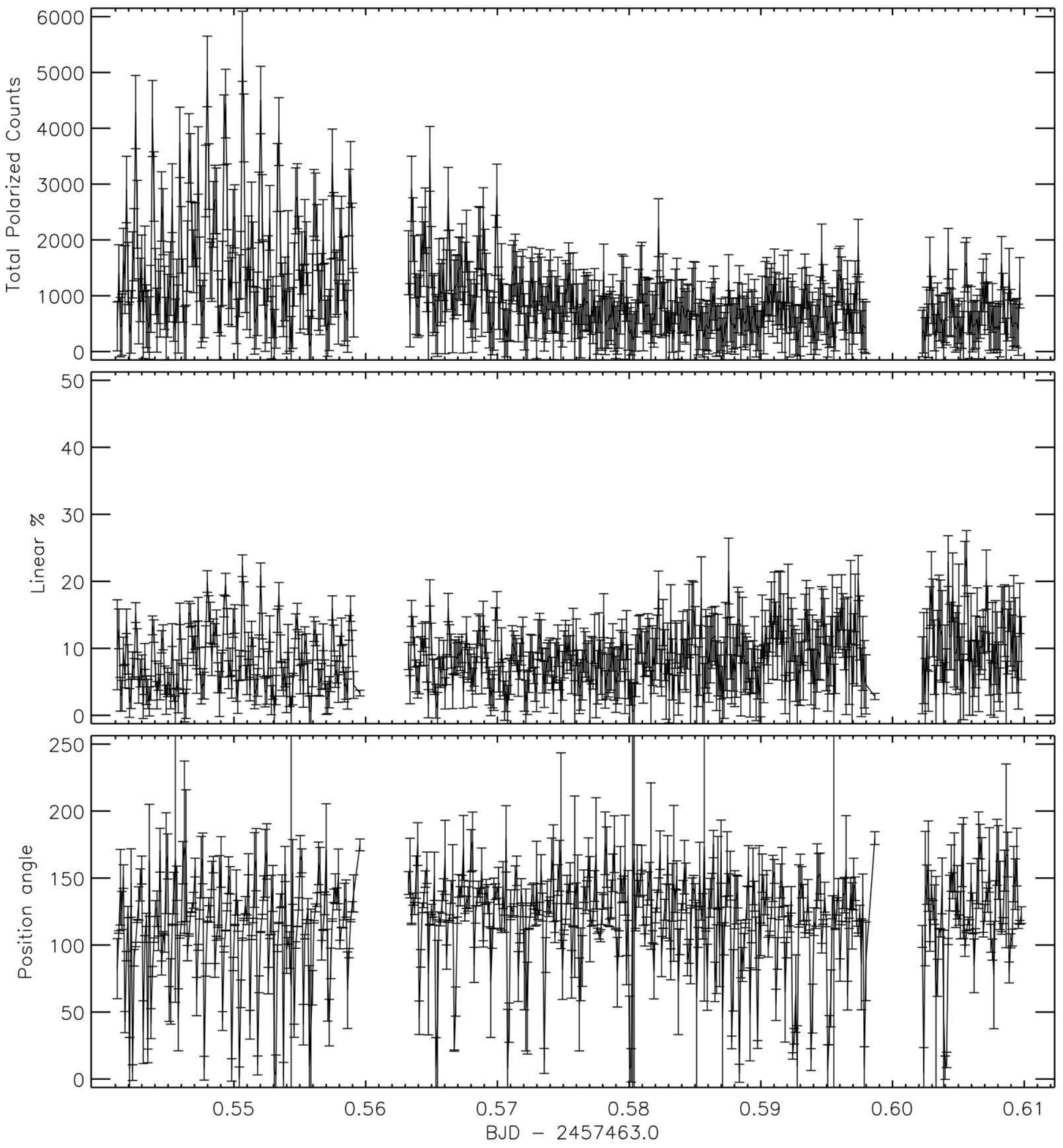}
\captionsetup{skip=-40pt}
		\caption{Time series polarimetry. \newline Broad band (340 -- 900 nm) photopolarimetry of AR Sco taken on 15 March 2016, in 10 s bins. The panels show, from the top,  the total polarized flux ({\it s}), degree of linear polarization ({\it p}) and position angle of linear polarization ({\it $\theta$}). The error bars are $\pm 1\sigma$.  The data cover orbital phase interval $\phi = 0.38-0.85$ and the gaps are when  background sky measurements were obtained.}
		\label{Fig. S5}
	\end{center}
\end{figure}

\newpage
\begin{figure}
	\begin{center}
	   \includegraphics[width=0.85\textwidth]{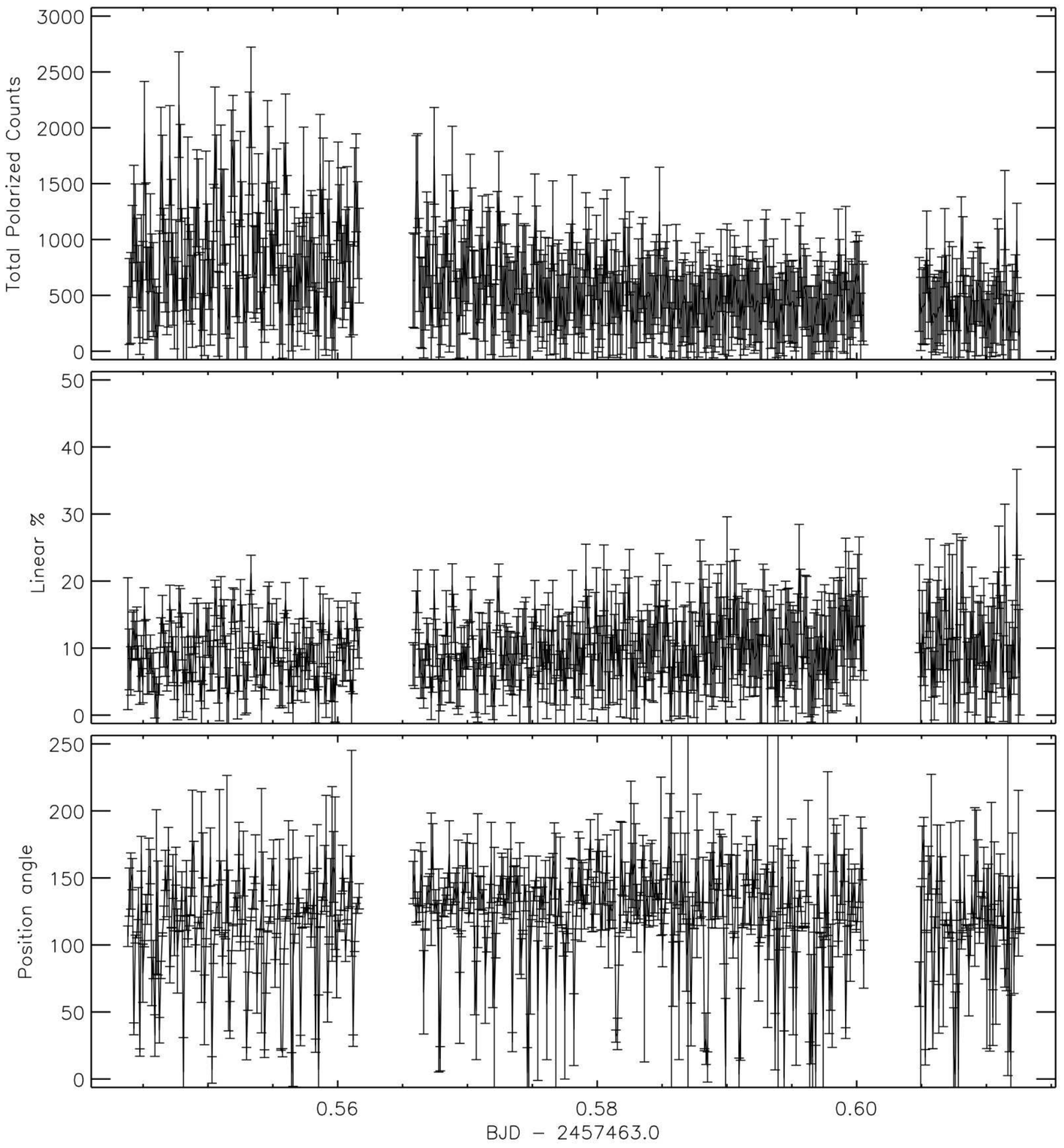}
\captionsetup{skip=-40pt}
		\caption{Time series polarimetry. \newline Red band (570 -- 900 nm) photopolarimetry of AR Sco taken on 15 March 2016, in 10 s bins. The panels show, from the top,  the total polarized flux ({\it s}), degree of linear polarization ({\it p}) and position angle of linear polarization ({\it $\theta$}). The error bars are $\pm 1\sigma$.  The data cover orbital phase interval $\phi = 0.40-0.86$ and the gaps are when background sky measurements were obtained. }
		\label{Fig. S6}
	\end{center}
\end{figure}

\newpage
\begin{figure}
	\begin{center}
	   \includegraphics[width=0.95\textwidth]{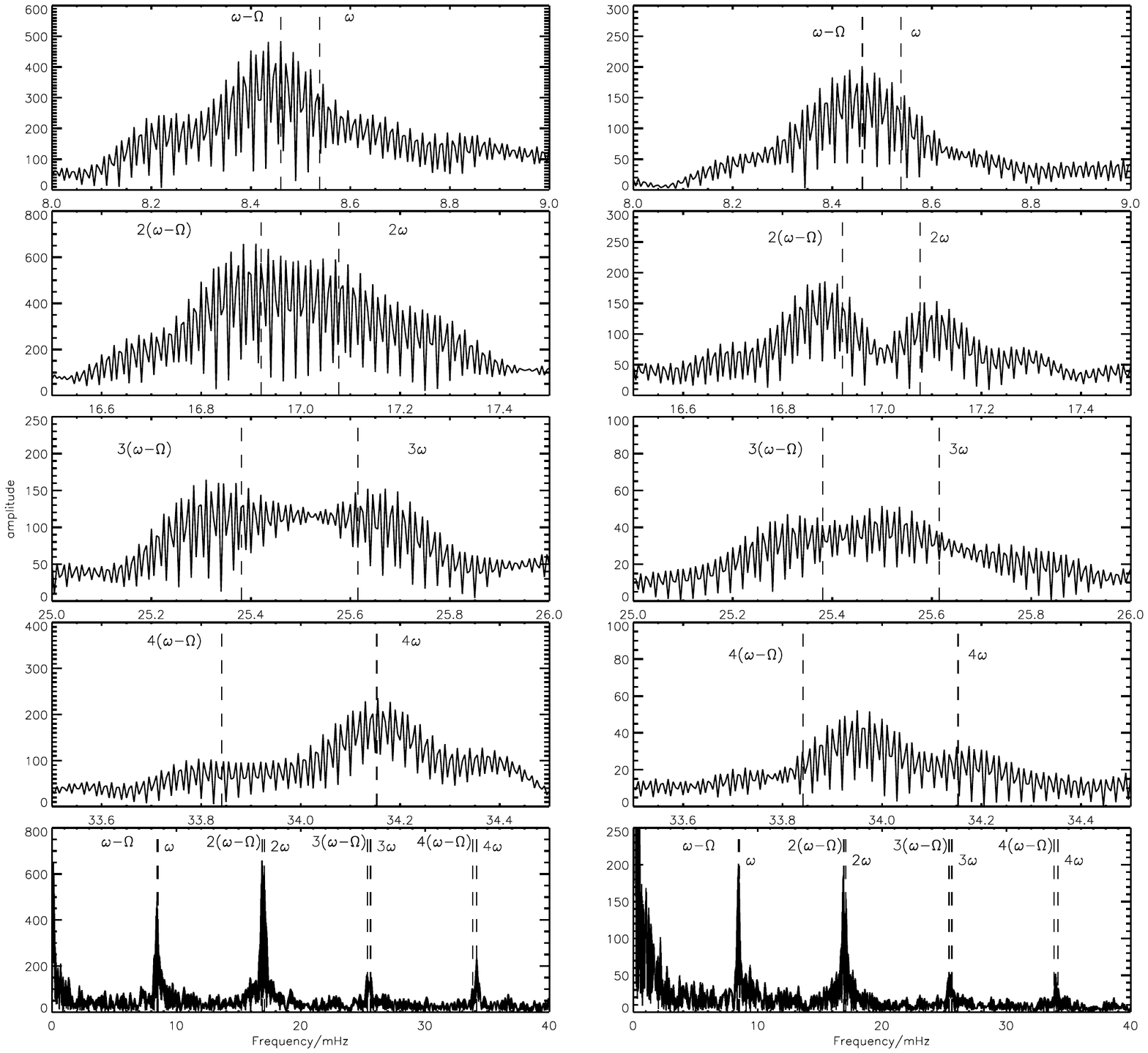}
\captionsetup{skip=-60pt}
		\caption{Polarization periodograms. \newline Amplitude spectra of the broadband polarized flux ({\it s}, left) and total flux ({\it I}, right)}
		\label{Fig. S7}
	\end{center}
\end{figure}

\newpage
\begin{figure}
	\begin{center}
	   \includegraphics[width=0.8\textwidth]{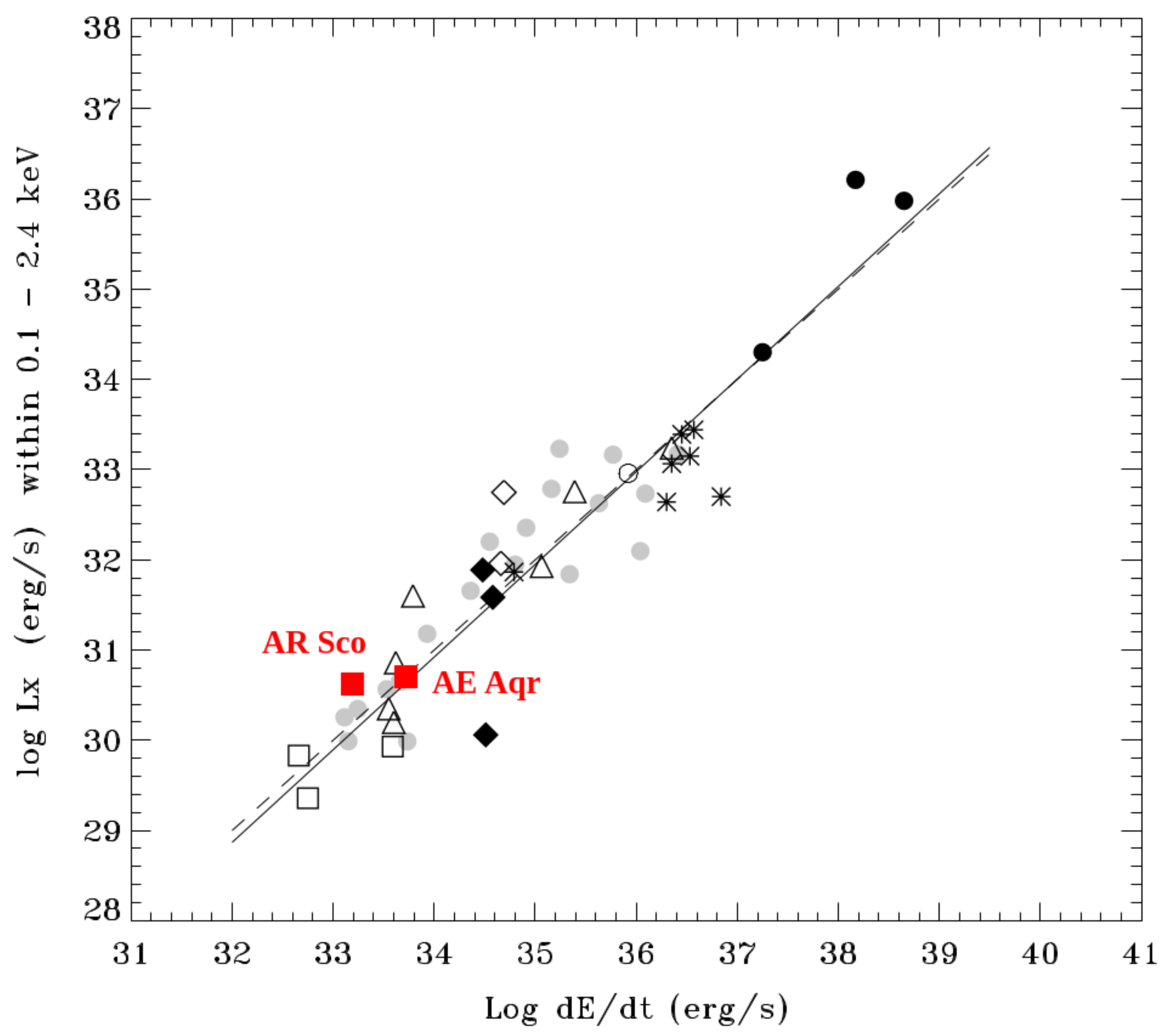}
\captionsetup{skip=30pt}
		\caption{Spin-down power vs. X-ray power. \newline The X-ray luminosity and spindown power for a sample of spin-powered pulsars. The straight line represents the $\alpha = 0.001$ ratio that is inherent to spin-powered pulsars and also the spin-powered white dwarfs in AE Aquarii and AR Sco (adapted from {\it 18\/}).}
		\label{Fig. S8}
	\end{center}
\end{figure}

\end{document}